\documentclass[11pt]{article}
\usepackage{latexsym, amssymb}
\title{Decay Rates and Probability Estimates for Massive Dirac Particles
in the Kerr-Newman Black Hole Geometry}
\date{July 2001 / January 2002}
\author{F.\ Finster, N.\ Kamran\thanks{Research supported by NSERC grant
\# RGPIN 105490-1998.}, J.\ Smoller\thanks{Research supported in
part by the NSF, Grant No.\ DMS-0103998.}, and S.-T.\
Yau\thanks{Research supported in part by the NSF, Grant No.\
33-585-7510-2-30.}}

\newtheorem{Def}{Def.}[section]
\newtheorem{Thm}[Def]{Theorem}

\newtheorem{Lemma}[Def]{Lemma}

\newtheorem{Corollary}[Def]{Corollary}
\newcommand{\Proof}{{\em{Proof. }}}
\newcommand{\QED}{\ \hfill $\FBox$ \\[1em]}
\newcommand{\spc}{\;\;\;\;\;\;\;\;\;\;}
\newcommand{\bra}{\mbox{$< \!\!$ \nolinebreak}}
\newcommand{\ket}{\mbox{\nolinebreak $>$}}
\newfont{\Bb}{msbm10 scaled 1095}
\newcommand{\C}{{\mbox{\Bb C}}}
\newcommand{\R}{{\mbox{\Bb R}}}

\newcommand{\sZ}{\mbox{\rm \bf \scriptsize Z}}
\newcommand{\sR}{\mbox{\rm \scriptsize I \hspace{-.8 em} R}}

\newcommand{\slsh}{\mbox{ \hspace{-1.1 em} $/$}}

\newcommand{\umin}{u_{\mbox{\scriptsize{min}}}}
\newcommand{\sumin}{u_{\mbox{\tiny{min}}}}
\newcommand{\xmin}{x_{\mbox{\scriptsize{min}}}}
\newcommand{\sxmin}{x_{\mbox{\tiny{min}}}}
\newcommand{\smin}{s_{\mbox{\scriptsize{min}}}}
\newcommand{\ssmin}{s_{\mbox{\tiny{min}}}}
\newcommand{\pmin}{\phi_{\mbox{\scriptsize{min}}}}

\newcommand{\FBox}{\rule{2mm}{2.25mm}}

\begin{document}

\include{epsf}
\maketitle

\begin{abstract}
The Cauchy problem is considered for the massive Dirac equation in
the non-extreme Kerr-Newman geometry, for smooth initial data with
compact support outside the event horizon and bounded angular
momentum. We prove that the Dirac wave function decays in
$L^\infty_{\mbox{\scriptsize{loc}}}$ at least at the rate
$t^{-5/6}$. For generic initial data, this rate of decay is sharp.
We derive a formula for the probability $p$ that the Dirac particle 
escapes to infinity. For various conditions on the initial data, we 
show that $p=0,1$ or $0<p<1$. The proofs are based on
a refined analysis of the Dirac propagator constructed
in~\cite{FKSY2}.
\end{abstract}

\tableofcontents

\section{Introduction}
\setcounter{equation}{0}
The Cauchy problem for the massive Dirac equation in the
non-extreme Kerr-Newman black hole geometry outside the event
horizon was recently studied~\cite{FKSY2}, and it was proved that
for initial data in $L^\infty_{\mbox{\scriptsize{loc}}}$ with
$L^2$ decay at infinity, the probability for the Dirac particle to
be located in any compact region of space tends to zero as $t
\rightarrow \infty$. This result shows that the Dirac particle
must eventually either disappear into the event
horizon or escape to infinity. The questions of the likelihood of
each of these possibilities and the rates of decay of the Dirac
wave function in a compact region of space
were left open. In the present
paper, we shall study these questions by means of a detailed analysis of the
integral representation of the Dirac propagator constructed
in~\cite{FKSY2}. This analysis will also give us some insight into
the physical mechanism which leads to the decay.

Recall that in Boyer-Lindquist coordinates $(t, r, \vartheta,
\varphi)$ with $r>0$, $0 \leq \vartheta \leq \pi$, $0 \leq \varphi
< 2\pi$, the Kerr-Newman metric is given by \cite{C}
\begin{eqnarray}
\lefteqn{ ds^2 \;=\; g_{jk}\:dx^j x^k } \nonumber \\ &=&
\frac{\Delta}{U} \:(dt \:-\: a \:\sin^2 \vartheta \:d\varphi)^2
\:-\: U \left( \frac{dr^2}{\Delta} + d\vartheta^2 \right) \:-\:
\frac{\sin^2 \vartheta}{U} \:(a \:dt \:-\: (r^2+a^2) \:d\varphi)^2
\spc \label{eq:0}
\end{eqnarray} with
\[ U(r, \vartheta) \;=\; r^2 + a^2 \:\cos^2 \vartheta \;,\spc
\Delta(r) \;=\; r^2 - 2 M r + a^2 + Q^2 \; , \] and the
electromagnetic potential is
\[ A_j \:dx^j \;=\; -\frac{Q \:r}{U} \:(dt \:-\:  a \:\sin^2 \vartheta \:
d\varphi) \; , \]
where $M$, $aM$ and $Q$ denote the mass, the angular momentum and
the charge of the black hole, respectively. Here $a$ and/or $Q$ are 
allowed to be zero, so that our results apply also to the Kerr, 
Reissner-Nordstr\"om, and Schwarzschild solutions. 
We shall restrict attention to the {\em{non-extreme case}} $M^2 >
a^2+Q^2$, in which case the function $\Delta$ has two distinct zeros,
\[ r_0 \;=\; M \:-\: \sqrt{M^2 - a^2 - Q^2} \spc {\mbox{and}} \spc r_1
\;=\; M \:+\: \sqrt{M^2 - a^2 - Q^2} \; , \] corresponding to the
Cauchy horizon and the event horizon, respectively.  We will here
consider only the region $r>r_1$ outside of the event horizon, and
thus $\Delta>0$.

Our starting point is the representation of the Dirac propagator for a
Dirac particle of mass $m$ and charge $e$ established
in~\cite[Thm.~3.6]{FKSY2}
\begin{equation}
\Psi(t,x) \;=\; \frac{1}{\pi} \: \sum_{k,n \in \sZ}
\int_{-\infty}^\infty d\omega \:e^{-i \omega t} \:\sum_{a,b=1}^2
t^{k \omega n}_{ab} \:\Psi^{k \omega n}_a(x) \:\bra \Psi^{k \omega
n}_b \:|\: \Psi_0 \ket \;.    \label{eq:1a}
\end{equation}
Here $\Psi_0$ is the initial data, $\omega$ is the energy, and $\bra .|.
\ket$ is a positive scalar product (see~\cite{FKSY2} for details).
The quantum number $k$ arises from the usual separation $\sim
\exp(-i(k+\frac{1}{2}))$ of the angular dependence around the axis
of symmetry, whereas $n$ labels the eigenvalues of generalized total
angular momentum in Chandrasekhar's separation of the Dirac equation
into ODEs~\cite{C2}. The $\Psi^{k \omega n}_a$ are solutions of the
Dirac equation and $t^{k \omega n}_{ab}$ are complex coefficients; they can
all be expressed in terms of the fundamental solutions to these
ODEs. We postpone the detailed formulas for $\Psi^{k \omega n}_a$ and $t^{k
\omega n}_{ab}$ to later sections, and here merely describe those
qualitative
properties of the wave functions $\Psi^{k \omega n}_a$ which are needed for
understanding our results. Near the event horizon, the $\Psi^{k \omega n}_a$
go over asymptotically to spherical waves. In the region $|\omega|>m$, the
solutions for $a=1$ are the {\em{incoming waves}}, i.e.\ asymptotically near
the event horizon they are waves moving towards the black hole. Conversely,
the solutions for $a=2$ are the {\em{outgoing waves}}, which near the event
horizon move outwards, away from the black hole. Asymptotically near
infinity, the $\Psi^{k \omega n}_a$, $|\omega|>m$, go over to spherical
waves.
In the region $|\omega|<m$, however, the fundamental solutions for $a=1,2$
near the event horizon are both linear combinations of incoming and outgoing
waves, taken in such a way that $\Psi^{k \omega n}_1$ and $\Psi^{k \omega
n}_2$ at infinity have exponential decay and growth, respectively.

For technical simplicity, we make the assumption that $\Psi_0$ is
{\em{smooth}} and {\em{compactly supported}} outside the event horizon.
We point out that, while the assumption of compact support is
physically reasonable at infinity, it is indeed restrictive with respect
to the behavior near the event horizon. Furthermore, we shall assume that the
{\em{angular momentum is bounded}} in the strict sense that there
exist constants $k_0$ and $n_0$ such that
\begin{equation}
\Psi(t,x) \;=\; \frac{1}{\pi} \: \sum_{|k| \leq k_0} \:\sum_{|n|
\leq n_0} \int_{-\infty}^\infty d\omega \:e^{-i \omega t} \left[
\sum_{a,b=1}^2 t^{k \omega n}_{ab} \:\Psi^{k \omega n}_a(x) \:\bra
\Psi^{k \omega n}_b \:|\: \Psi_0 \ket \right] \;.    \label{eq:1b}
\end{equation}
We expect that the rate of decay is the same if an
infinite number of angular modes are present. Namely, away from the
event horizon, modes with large angular momentum feel strong centrifugal
forces and should therefore be quickly driven out to infinity,
whereas the behavior near the event horizon is independent of
the angular momentum.
However, it seems a very delicate problem to rigorously establish decay
rates without the assumption~(\ref{eq:1b}), because this would make
it necessary to control the dependence of our estimates on $k$ and $n$.
Finally, we assume that the charge of the black
hole is so small that the gravitational attraction is the dominant force
at a large distance from the black hole. More precisely, we shall assume
throughout this paper that
\begin{equation}
m\:M \;>\; |e\:Q| \;.\label{eq:cc}
\end{equation}

We now state our main results and discuss them afterwards.
\begin{Thm}[Decay Rates] \label{thm1}
Consider the Cauchy problem
\[ (i \gamma^j D_j - m)\: \Psi(t,x) \;=\; 0\;,\spc
\Psi(0,x) \;=\; \Psi_0(x) \] for the Dirac equation in the
non-extreme Kerr-Newman black hole geometry with small
charge~(\ref{eq:cc}). Assume that the Cauchy data $\Psi_0$ is
smooth with compact support outside the event horizon and has
bounded angular momentum~(\ref{eq:1b}).
\begin{description}
\item[(i)] If for any $k$ and $n$,
\begin{equation}
\limsup_{\omega \searrow m} |\bra \Psi^{k \omega n}_2 \:|\: \Psi_0
\ket | \;\neq\; 0 \spc {\mbox{or}}\spc
\limsup_{\omega \nearrow -m} |\bra \Psi^{k \omega n}_2 \:|\: \Psi_0 \ket|
\;\neq\; 0\;, \label{eq:1as1}
\end{equation}
then for large $t$,
\begin{equation}
|\Psi(t,x)| \;=\; c\: t^{-\frac{5}{6}} + {\cal{O}}(t^{-\frac{5}{6}
-\varepsilon}) \;, \label{eq:1decay}
\end{equation}
with $c=c(x) \neq 0$ and any $\varepsilon <\frac{1}{30}$.
\item[(ii)] If for all $k, n$ and $a=1,2$,
\[ \bra \Psi^{k \omega n}_a \:|\: \Psi_0 \ket \;=\; 0 \]
for all $\omega$ in a neighborhood of $\pm m$, then $|\Psi(t,x)|$
has rapid decay in $t$ (i.e.\ for 
any fixed $x$, $\Psi(t,x)$ decays in $t$ faster than polynomially).
\end{description}
\end{Thm}

\begin{Thm}[Probability Estimates] \label{thm2}
Consider the Cauchy problem as in Theorem \ref{thm1}, with initial
data $\Psi_0$ normalized by $\bra \Psi_0 \:|\: \Psi_0 \ket = 1$. Let $p$ be
the probability for the Dirac particle to escape to infinity,
defined for any $R>r_1$ by
\begin{equation}
p \;=\; \lim_{t \rightarrow \infty} \int_{\{R < r <
\infty \}} (\overline{\Psi} \gamma^j \Psi)(t,x) \:\nu_j
\; d\mu \;, \label{eq:1pdef}
\end{equation}
where $\nu$ denotes the future
directed normal to the hypersurface $t={\mbox{const}}$ and $d\mu$
denotes the induced invariant measure on that hypersurface. Then $p$ is
given by
\begin{equation}
p \;=\; \frac{1}{\pi} \: \sum_{|k| \leq k_0} \:\sum_{|n| \leq n_0}
\int_{\sR \setminus [-m,m]} d\omega \left( \frac{1}{2} -
2 \:|t^{k \omega n}_{12}|^2 \right) \left| \bra \Psi^{k \omega n}_2 \:|\:
\Psi_0 \ket \right|^2 \;. \label{eq:1prob}
\end{equation}
Accordingly, $1-p$ gives the probability that the Dirac particles 
disappears into the event horizon. Furthermore,
\begin{description}
\item[(i)] Suppose that the outgoing energy distribution for $|\omega|>m$ is

non-zero, i.e.
\[ \bra \Psi^{k \omega n}_2 \:|\: \Psi_0 \ket \neq 0 \]
for some $\omega$ with $|\omega|>m$. Then $p>0$.
\item[(ii)] If the energy distribution of the Cauchy data has a 
non-zero contribution in the interval $[-m,m]$, then $p<1$.
\item[(iii)] If the energy distribution of the Cauchy data is 
supported in $[-m,m]$, then $p=0$.
\item[(iv)] If~(\ref{eq:1as1}) holds, then $0<p<1$.
\end{description}
\end{Thm}

\begin{figure}[tb]
    \epsfxsize=16cm
    \centerline{\epsfbox{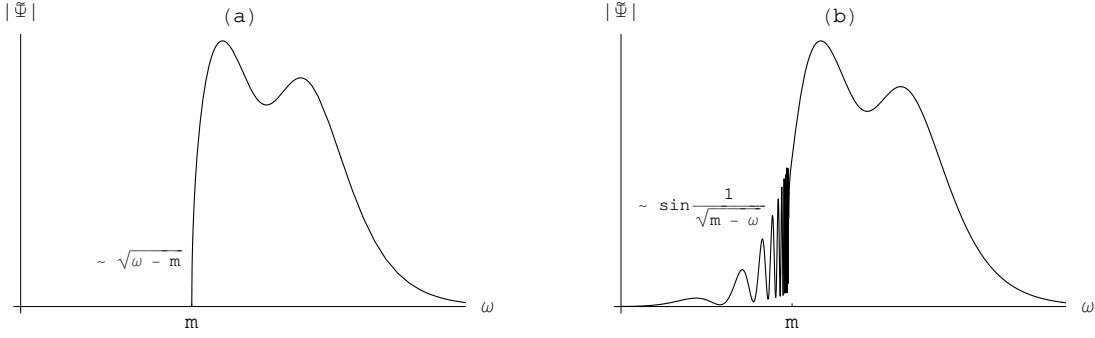}}
    \caption{Typical plot for $\tilde{\Psi}$ in Minkowski space (a) and
in the Kerr-Newman black hole geometry (b).}
    \label{fig1}
\end{figure}

The decay rate of $t^{-\frac{5}{6}}$ obtained in
Theorem~\ref{thm1} quantifies the effect of the black hole's
gravitational attraction on the long-time behavior of massive
Dirac particles. Before discussing this effect in detail, it is
instructive to recall the derivation of the decay rates in
Minkowski space. We denote the plane-wave solutions of the Dirac
equation by $\Psi_{\vec{k} s \epsilon}$, where $\vec{k}$ is
momentum, $\epsilon=\pm1$ is the sign of energy, and $s=\pm$
refers to the two spin orientations. The plane-wave solutions are
normalized according to
\[ (\Psi_{\vec{k} s \epsilon} \:|\: \Psi_{\vec{k}' s' \epsilon'})
\;=\; \delta(\vec{k}-\vec{k}')\: \delta_{s s'}\: \delta_{\epsilon
\epsilon'} \;, \] where $(.|.)$ is the usual spatial scalar
product
\[ (\Psi \:|\: \Phi)(t) \;=\; \int \overline{\Psi(t, \vec{x})}\: \gamma^0\:
\Phi(t, \vec{x})\: d\vec{x}\;. \] The Dirac propagator is obtained
by decomposing the initial data into the plane-wave solutions,
\[ \Psi(t, \vec{x}) \;=\; \sum_{s, \epsilon} \int d\vec{k}\;
\Psi_{\vec{k} s \epsilon}(t, \vec{x}) \: (\Psi_{\vec{k} s
\epsilon}(t=0) \:|\: \Psi_0)\;, \] and a straightforward
calculation using the explicit form of the plane-wave solutions
yields that
\begin{equation}
\Psi(t, \vec{x}) \;=\; 2 \pi \int \frac{d^4k}{(2 \pi)^4}\: (k
\slsh + m)\: \delta(k^2-m^2)\: \Gamma(\omega)\:
e^{-ikx}\:\gamma^0\: \hat{\Psi}_0(\vec{k})\;, \label{eq:I1}
\end{equation}
where $\hat{\Psi}_0(\vec{k})$ is the Fourier transform of
$\Psi_0(\vec{x})$ (and as usual $x=(t, \vec{x})$, $k=(\omega,
\vec{k})$, $k \slsh = k_j \gamma^j$, and $\Gamma$ is the step function
$\Gamma(x) = {\mbox{sgn}}(x)$).
Let us assume for simplicity that the initial
data is a Schwartz function.  We write~(\ref{eq:I1}) as a Fourier
integral in $\omega$,
\begin{equation}
\Psi(t, \vec{x}) \;=\; \int_{-\infty}^\infty \tilde{\Psi}(\omega,
\vec{x}) \: e^{-i \omega t}\: d\omega \label{eq:I2}
\end{equation}
where
\[ \tilde{\Psi}(\omega, \vec{x}) \;=\; \int \frac{d\vec{k}}{(2 \pi)^3}\:
(k \slsh+m)\: \delta(\omega^2 - |\vec{k}|^2 - m^2)\:
\Gamma(\omega)\: e^{i \vec{k} \vec{x}}\: \gamma^0\:
\hat{\Psi}_0(\vec{k})\;. \] We consider the $\omega$-dependence of
$\tilde{\Psi}$ for fixed $\vec{x}$. The $\delta$-distribution
gives a contribution to the momentum integral only for $\vec{k}$
on the sphere $|\vec{k}|^2=\omega^2-m^2$. Thus
$\tilde{\Psi}(\omega, \vec{x})$ vanishes for $|\omega|<m$ and has
rapid decay at infinity. Furthermore, $\tilde{\Psi}$ is clearly
smooth in the region $|\omega|>m$. For $|\omega|$ near $m$,
$\tilde{\Psi}$ has the expansion
\begin{eqnarray*}
\tilde{\Psi}(\omega, \vec{x}) &=& \int_0^\infty \frac{k^2\:dk}{4
\pi^2}\: (\omega \gamma^0 + m)\: \delta(\omega^2 - k^2 - m^2)\:
\Gamma(\omega)\:
\gamma^0\: \hat{\Psi}_0(0)\; (1+{\cal{O}}(k)) \\
&=& \frac{\Gamma(\omega)}{8 \pi^2}\: (\omega + m \gamma^0)\:
\hat{\Psi}_0(0)\: \sqrt{\omega^2 - m^2}\:+\:
{\cal{O}}(\omega^2-m^2)\;.
\end{eqnarray*}
A typical plot of $|\tilde{\Psi}(\omega, \vec{x})|$ is shown in
Figure~\ref{fig1}(a). If $\hat{\Psi}_0$ vanishes in a neighborhood
of $\vec{k}=0$, then $\tilde{\Psi}(.,\vec{x})$ is a Schwartz
function, and thus its Fourier transform~(\ref{eq:I2}) has rapid
decay. This is the analogue of Case~{\bf{(ii)}} of Theorem~\ref{thm1}.
However, if $\tilde{\Psi}_0(0) \neq 0$, the decay rate is
determined by the square root behavior of $\tilde{\Psi}$ for
$|\omega|$ near $m$. A change of variables gives that for
any test function $\eta$ which is supported in a neighborhood of
the origin,
\[ \int_m^\infty \sqrt{\omega - m}\: \eta(\omega-m)\: e^{-i \omega t}\:
d\omega \;=\; e^{-imt}\: t^{-\frac{3}{2}}\: \int_0^\infty
\sqrt{u}\: \eta\!\left( \frac{u}{t} \right) e^{-iu}\:du\;. \]
An integration-by-parts argument shows that the last integral is
bounded uniformly in $t$, and is non-zero for large $t$ if
$\eta(0) \neq 0$. From this we conclude that in Minkowski space,
$|\Psi(t, \vec{x})|$ decays polynomially at the rate $t^{-\frac{3}{2}}$.

We now proceed with a more detailed discussion of our results,
beginning with the rates of decay obtained in Theorem~\ref{thm1}.
Naively speaking, a massive Dirac particle behaves near the
event horizon similar to a massless particle, i.e.\ like a
solution of the wave equation. In Minkowski space, solutions of the 
wave equation decay rapidly in time according to the Huygens principle. On
the other hand, at large distance from the black hole the
solutions should behave like those of the massive Dirac equation
in Minkowski space, which decay at the rate $t^{-\frac{3}{2}}$.
It is thus tempting to expect that the solutions of the massive
Dirac equation in the Kerr-Newman black hole geometry should decay
at a rate which ``interpolates'' between the behavior of a massive
particle in Minkowski space and that of a massless particle, and
should thus decay at a rate no slower than $t^{-\frac{3}{2}}$.
However, Theorem~\ref{thm1} shows that this naive picture is
incorrect, since the rate of decay we have established for a
massive Dirac particle in the Kerr-Newman black hole geometry is
actually slower than that of a massive particle in Minkowski
space. Thus the gravitational field of the black hole affects the
behavior of massive Dirac particles in a more subtle way. One can
understand this fact by comparing the plots in Figure~\ref{fig1}, which
give typical examples for the energy distribution of the Dirac
wave function in Minkowski space and in the Kerr-Newman geometry.
One sees that in the Kerr-Newman geometry, there is a contribution
to the energy distribution for $|\omega|<m$, which oscillates
infinitely fast as $\omega$ approaches $m$. When taking the
Fourier transform, these oscillations lead to the decay rate
$t^{-\frac{5}{6}}$ given Theorem~\ref{thm1} (see the
rigorous saddle point argument in Lemma~\ref{lemma42}).

The oscillations in the energy distribution in Figure~\ref{fig1}(b) 
are a consequence of the field behavior near spatial infinity. On
a qualitative level, they can already be understood in
Newtonian gravity and the semi-classical approximation. Namely,
in the Newtonian limit of General Relativity, the momentum $\vec{k}$ of a
relativistic particle is related to its energy $\omega$ by
\[ |\vec{k}|^2 \;=\; \left( \omega + \frac{m M}{r} \right)^2 -
m^2\;. \] Thus the particle has positive momentum even if
$\omega<m$, provided that the Newtonian potential is large enough,
$\frac{mM}{r} > m-\omega$. This means in the semi-classical
approximation that the wave function $\Psi(r)$ has an oscillatory
behavior near the black hole,
\[ \Psi(r) \;\sim\; \exp \left( \pm i \int^r \vec{k} \:d\vec{s}
\right) \spc{\mbox{for}}\spc r< R \equiv \frac{mM}{m-\omega}\;,
\] and will fall off exponentially for $r>R$. As a consequence,
the fundamental solutions $\Psi^{k \omega n}_a$ for $|\omega|<m$
involve phase factors $\sim \exp (\pm i \int^{R} \vec{k}\:
d\vec{s})$. In the limit $\omega \nearrow m$, $R \rightarrow
\infty$, leading to infinitely fast oscillations in our integral
representation. This simple argument even gives the correct
quantitative behavior of the phases $\sim
(m-\omega)^{-\frac{1}{2}}$.

The fact that the decay rate in the presence of a black hole is
slower than in Minkowski space has the following direct physical
interpretation. One can view the gravitational attraction of the
black hole and the tendency of quantum mechanical wave functions
to spread out in space as competing with each other over time. The
component of the wave function for $\omega$ near $m$ and $\omega<m$,
which is responsible for the decay rate $t^{-\frac{5}{6}}$, has not enough
energy to propagate out to infinity. But since it is an outgoing wave near
the event horizon (note that in~(\ref{eq:1as1}) the fundamental solutions
$\Psi^{k
\omega n}_a$ enter only for $a=2$), it is driven outwards and resists the
gravitational attraction for a long time before it will eventually be drawn
into the black hole. As a result, the Dirac particle stays in any compact
region of space longer than it would in Minkowski space, and thus the rate
of decay of the wave function is slower.

According to this interpretation, our decay rates are a
consequence of the far-field behavior of the black hole.
Similar to the ``power law tails'' in the massless case
(see~\cite{P}), our effect
can be understood as a ``backscattering'' of the outgoing wave from
the long-range potential, but clearly the rest mass drastically changes
the behavior of the wave near infinity. We expect that result for the
decay rates should be valid even in a more general
setting, independent of the details of the local geometry near the event
horizon. Furthermore, the decay rates should be independent of the spin.
This view is supported by~\cite{KT1, KT2}, who obtained the rate
$t^{-\frac{5}{6}}$ for massive scalar fields in a spherically symmetric
geometry using asymptotic expansions of the Green's functions.

Theorem~\ref{thm2} gives a precise formula for the probability that 
the Dirac particle either disappears into the black hole or escapes to 
infinity. In cases {\bf{(i)}}--{\bf{(iv)}} we give sufficient conditions for
these  probabilities to occur. These results are consistent with the general
behavior of quantum mechanical
particles in the presence of a potential barrier and can be
thought of as a tunnelling effect. 
In case {\bf{(iii)}}, the particle does not have 
enough energy to escape to infinity. Thinking again in terms of a 
tunnelling effect, the Dirac particle cannot tunnel to infinity 
because the potential barrier (which has finite height $m$) has 
infinite width. Finally, one might ask whether $p=1$ can occur; i.e., 
that the particles escapes to infinity with probability one. This is 
indeed the case for very special initial data, whose energy 
distribution is supported outside the interval $[-m,m]$ (see
Corollary~\ref{cor91} below).

We conclude by remarking that a number of significant results are
known for the long-term behavior of {\em{massless}} fields in
black hole geometries. These results do not capture our effect,
which is intimately related to the presence of a mass gap in the
energy spectrum. Price~\cite{P} discussed the rates of decay
of massless fields in the Schwarzschild background for special choices of
initial data. His decay rates depend on the angular
momentum and are faster than the ones we have derived. A rigorous
proof of the boundedness of the solutions of the wave equation in
the Schwarzschild geometry has been given by Kay and Wald~\cite{KW}.
Beyer pursues an approach using $C^0$-semigroup theory, which also
applies to the Kerr metric and the massive case~\cite{B}.
An important contribution to the long-time
behavior of gravitational perturbations of the Kerr metric has
been given by Whiting~\cite{W}.

\section{The Long-Time Dynamics under a Spectral Condition}
\setcounter{equation}{0}
We begin the analysis with the case when the energy distribution
of the Cauchy data is zero in a neighborhood of $\omega = \pm m$.
The following theorem is an equivalent formulation of
Theorem~\ref{thm1}{\bf{(ii)}}.
\begin{Thm} \label{thm21}
Consider the Cauchy problem
\[ (i \gamma^j D_j - m) \: \Psi(t,x)
\;=\; 0 \;\;\;,\spc \Psi(0,x) \;=\; \Psi_0(x) \]
for smooth initial data with compact support outside the event
horizon.  Assume that angular momentum is bounded and that the energy
is supported away from $\omega = \pm m$, i.e.
\begin{equation}
\Psi_0 \;=\; \frac{1}{\pi} \: \sum_{|k| \leq k_0} \:\sum_{|n| \leq
n_0} \left( \int_{-\infty}^{-m-\varepsilon} \!+
\int_{-m+\varepsilon}^{m-\varepsilon} \!+
\int_{m+\varepsilon}^\infty \right) d\omega \:\sum_{a,b=1}^2 t^{k
\omega n}_{ab} \:\Psi^{k \omega n}_a \:\bra \Psi^{k \omega n}_b
\:|\: \Psi_0 \ket    \label{eq:1c}
\end{equation}
for suitable $\varepsilon>0$. Then for all $x$, $\Psi(t,x)$ has
rapid decay in $t$.
\end{Thm}

Before giving the proof, we recall a few basic formulas
from~\cite{FKSY2}. The separation ansatz for the fundamental solutions
$\Psi^{k \omega n}_a$ is
\begin{equation}
\Psi^{k \omega n}_a(t, r, \vartheta, \varphi) \;=\;
e^{-i \omega t}\: e^{-i(k+\frac{1}{2})\varphi} \left(
\begin{array}{c}
X^{k \omega n}_-(r)\: Y^{k \omega n}_-(\vartheta) \\
X^{k \omega n}_+(r)\: Y^{k \omega n}_+(\vartheta) \\
X^{k \omega n}_+(r)\: Y^{k \omega n}_-(\vartheta) \\
X^{k \omega n}_-(r)\: Y^{k \omega n}_+(\vartheta)
\end{array} \right) , \label{eq:323}
\end{equation}
where $X=(X_+, X_-)$ and $Y=(Y_+, Y_-)$ are the radial and angular
components, respectively. The radial part $X(u)$ is a solution of the radial
Dirac equation~\cite[eqn (3.7)]{FKSY2}
\begin{eqnarray}
\left[ \frac{d}{du} + i \Omega(u) \left( \begin{array}{cc} 1 & 0
\\ 0 & -1 \end{array} \right) \right] X
\;=\; \frac{\sqrt{\Delta}}{r^2+a^2} \:\left( \begin{array}{cc} 0 &
imr-\lambda \\
-imr-\lambda & 0 \end{array} \right) X \;, \label{eq:a}
\end{eqnarray}
where
\[ \Omega(u) \;=\; \omega+\frac{(k+\frac{1}{2}) \:a+eQr}{r^2+a^2}
\;\;\;,\spc \Delta \;=\; r^2-2Mr + a^2 + Q^2 \;, \] $\lambda$ is
the angular eigenvalue (which depends smoothly on $\omega$), and
$u \in (-\infty, \infty)$ is related to the radius by
\begin{equation}
\frac{du}{dr} \;=\; \frac{r^2+a^2}{\Delta}\;.
\label{eq:2ut}
\end{equation}

To analyze $X$ in the asymptotic region $u \rightarrow -\infty$,
one employs for $X$ the ansatz
\begin{equation}
X(u) \;=\; \left( \begin{array}{c} e^{-i \Omega_0 u} \:f^+(u) \\
e^{i \Omega_0 u} \:f^-(u) \end{array} \right) \label{eq:36}
\end{equation}
and obtains for $f$ the equation
\begin{eqnarray}
\frac{d}{du} \:f &=& \left[ i (\Omega_0 - \Omega(u)) \left(
\begin{array}{cc}
1 & 0 \\ 0 & -1 \end{array} \right) \right. \nonumber \\
&& \left. \;\;+\: \frac{\sqrt{\Delta}}{r^2+a^2}
\:\left( \begin{array}{cc} 0 & e^{-2i \Omega u} (imr-\lambda) \\
e^{2i \Omega u} (-imr-\lambda) & 0 \end{array} \right) \right] f
\;\;\;. \label{eq:1d}
\end{eqnarray}
Standard Gronwall estimates yield that the fundamental solutions
of~(\ref{eq:a}) have the asymptotic form~\cite[Lemma~3.1]{FKSY2}
\begin{equation}
X_a(u) \;=\; \left( \begin{array}{c} e^{-i \Omega_0 u} \: f^+_{0\:a} \\
e^{i \Omega_0 u} \: f^-_{0\:a} \end{array} \right) \:+\: R_0(u)
\;, \label{eq:c}
\end{equation}
where $|R_0(u)| \leq c\:\exp(du)$ for suitable constants $c,d>0$
and
\begin{equation}
f_{0\:1} \;=\; \left( \begin{array}{c} 1 \\ 0 \end{array} \right)
\;, \spc f_{0\:2} \;=\; \left( \begin{array}{c} 0 \\ 1 \end{array}
\right) \;. \label{eq:d}
\end{equation}
In the asymptotic region $u \rightarrow \infty$, one transforms
the spinor basis with a matrix $B(u)$ such that the matrix
potential in~(\ref{eq:a}) becomes equal to the diagonal matrix $-i
\Omega(u) \sigma^3$ ($\sigma^j$ are the Pauli matrices).
One must distinguish between the two cases
$|\omega|<m$ and $|\omega|>m$. In the first case, $\Omega(u)$ is imaginary
for large $u$, and thus there are two fundamental solutions
$X_1$ and $X_2$ with exponential decay and growth, respectively, and we
normalize them such that
\begin{equation}
\lim_{u \rightarrow -\infty} |X(u)| \;=\; 1 \;. \label{eq:2norm}
\end{equation}
In the case $|\omega|>m$, $\Omega(u)$ is real for all $u$. The ansatz
\begin{equation}
X \;=\; B \:\left( \begin{array}{c} e^{-i \Phi} \:f^+(u) \\ e^{i
\Phi} \:f^-(u)
\end{array} \right) \spc {\mbox{with}} \spc \Phi'(u)  \;=\; \Omega(u)
\label{eq:3C}
\end{equation}
gives the differential equation
\begin{equation}
\frac{d}{du} f \;=\; M(u) \:f \spc{\mbox{with}}\spc |M(u)| \leq
\frac{c}{u^2} \;,\label{eq:1e}
\end{equation}
which can again be controlled by Gronwall estimates. Thus one
obtains the asymptotic formula~\cite[Lemma~3.5]{FKSY2}
\begin{equation}
X_a(u) \;=\; \left( \begin{array}{cc} \cosh \Theta & \sinh \Theta \\
\sinh \Theta & \cosh \Theta \end{array} \right) \left(
\begin{array}{c} e^{-i \Phi(u)} \: f^+_{\infty\:a} \\
e^{i \Phi(u)} \: f^-_{\infty\:a} \end{array} \right) \:+\:
R_\infty(u) \;, \label{eq:b}
\end{equation}
where $|R_\infty| \leq c/u$ for suitable $c>0$ and
\begin{equation}
\Theta \;=\; \frac{1}{4} \:\log \left( \frac{\omega+m}{\omega-m}
\right)\;\;,\;\;\;\;\; \Phi \;=\; \Gamma(\omega) \left(
\sqrt{\omega^2-m^2} \:u \:+\: \frac{\omega e Q + M m^2}
{\sqrt{\omega^2-m^2}} \:\log u \right) . \label{eq:14a}
\end{equation}
The complex factors $f^\pm_{\infty\:a}$ in~(\ref{eq:b}) are the
so-called {\em{transmission coefficients}}. Furthermore, we
introduced the functions $t_a(\alpha)$, $0 \leq \alpha \leq 2
\pi$, in terms of the transmission coefficients
by~\cite[eqn~(3.47)]{FKSY2}
\begin{equation}
t_1(\alpha) \;=\; f_{\infty\: 2}^+ \:e^{-i \alpha} \:-\:
f_{\infty\: 2}^- \:e^{i \alpha} \;\;\;,\spc t_2(\alpha) \;=\;
-f_{\infty\:1}^+ \:e^{-i \alpha} \:+\: f_{\infty\:1}^- \:e^{i
\alpha} \;. \label{eq:j}
\end{equation}
Finally, the coefficients $(t_{ab})_{a,b=1,2}$ are given by
\begin{equation}
t_{ab} \;=\; \left\{ \begin{array}{lc}
\delta_{a1}\: \delta_{b1} & {\mbox{if $|\omega| \leq m$}} \\[.3em]
\displaystyle \frac{1}{2 \pi} \:\int_0^{2 \pi}
\frac{t_a \:\overline{t_b}}{|t_1|^2 + |t_2|^2} \:d\alpha &
{\mbox{if $|\omega|>m$.}} \end{array} \right. \label{eq:k}
\end{equation}
{{\em{Proof of Theorem~\ref{thm21}. }}} Since~(\ref{eq:1c})
contains only finite sums, we can fix $k, n$ and consider one
summand. The coefficients in the differential
equation~(\ref{eq:1d}) are smooth in $\omega$, and their
$\omega$-derivatives are integrable on the half-lines $u \in
(-\infty, u_0]$ for $u_0$ sufficiently small. Apart from the
singularities at $\omega=\pm m$, the same is true for the
differential equation~(\ref{eq:1e}) for $u$ on the half line
$[u_1, \infty)$ and $u_1$ sufficiently large. Since furthermore
the ansatz~(\ref{eq:36}) is smooth in $\omega \in \R \setminus
((-m-\varepsilon, -m+\varepsilon) \cup (m-\varepsilon,
m+\varepsilon))$ and~(\ref{eq:3C}) is smooth in $\omega \in
(-\infty, -m-\varepsilon] \cup [m+\varepsilon, \infty)$, and also
the coefficients of the differential equation~(\ref{eq:a}) depend
smoothly on $\omega$ in the bounded interval $u \in [u_0, u_1]$,
we conclude that the fundamental solutions $\Psi^{k_0 \omega
n_0}_a(x)$ and the transmission coefficients $f^{k_0 \omega
n_0}_a$ depend smoothly on $\omega \in \R \setminus
((-m-\varepsilon, -m+\varepsilon) \cup (m-\varepsilon,
m+\varepsilon))$. Hence the integrand
in~(\ref{eq:1c}) is a smooth function in $\omega$ (which vanishes
for $|\omega|-m<\varepsilon$). Since $\Psi_0$ has compact support and
the fundamental solutions for $\omega \rightarrow \infty$ go over to
plane waves, it is clear that the $\omega$-derivatives of
the integrand in~(\ref{eq:1c}) are all integrable. It follows that
the Fourier integral~(\ref{eq:1a}) has rapid decay.
\QED
According to this theorem and using linearity, it remains to
analyze the energy distribution in a neighborhood of $\omega = \pm
m$. Since all constructions and estimates are similar for positive
and negative $\omega$, we can in what follows restrict attention
to a neighborhood of $\omega=+m$.

\section{Decay Rates of Fourier Transforms -- Basic Considerations}
\setcounter{equation}{0}
In this section, we derive estimates of some elementary Fourier
integrals. Our decay rate $t^{-\frac{5}{6}}$ ultimately comes from
Lemma~\ref{lemma42}. We always denote by $\varepsilon$ a parameter in the
range $0<\varepsilon<\frac{1}{30}$.

\begin{Lemma} \label{lemma40}
Let $g \in L^\infty(\R) \cap C^1((0, \infty))$ with compact
support and assume that for a suitable constant $C$,
\begin{equation}
|g'(\alpha)| \;\leq\; \frac{C}{\alpha} \spc {\mbox{for all
$\alpha>0$.}} \label{eq:4a}
\end{equation}
Then there is a constant $c=c(g)$ such that for all $t>0$,
\[ \left| \int_0^\infty e^{i \alpha t}\: g(\alpha)\: d\alpha \right|
\;\leq\;
c\: t^{-\frac{5}{6}-\varepsilon}\;. \]
\end{Lemma}
{\Proof} Assume that ${\mbox{supp}}\: g \subset [-L, L]$. For
given $\delta>0$, we split up the integral as
\[ \int_0^\infty e^{i \alpha t}\: g(\alpha)\: d\alpha \;=\;
\int_0^\delta e^{i \alpha t}\: g(\alpha)\: d\alpha \:+\:
\int_\delta^\infty e^{i \alpha t}\: g(\alpha)\: d\alpha \;. \] The
first term can be estimated by
\[ \left| \int_0^\delta e^{i \alpha t}\: g(\alpha)\: d\alpha \right|
\;\leq\;
c_1\: \delta \] with $c_1=\sup |g|$. In the second term, we
integrate by parts,
\begin{eqnarray*}
\lefteqn{ \int_\delta^\infty e^{i \alpha t}\: g(\alpha)\: d\alpha
\;=\; \frac{1}{it}\: \int_\delta^\infty \left( \frac{d}{d\alpha}\:
e^{i \alpha t}\right) g(\alpha) \: d\alpha} \\
&=&-\frac{1}{it}\: e^{i \delta t}\: g(\delta) \:-\: \frac{1}{it}\:
\int_\delta^\infty e^{i \alpha t} g'(\alpha) \: d\alpha \;,
\end{eqnarray*}
and estimate using~(\ref{eq:4a}),
\[ \left| \int_\delta^\infty e^{i \alpha t}\: g(\alpha)\: d\alpha \right|
\;\leq\; \frac{c_1}{t}\:+\: \frac{C}{t}\: (\log L - \log \delta)
\;. \] We choose $\delta = t^{-\frac{5}{6}-\varepsilon}$ to
conclude that
\[ \left| \int_0^\infty e^{i \alpha t}\: g(\alpha)\: d\alpha \right|
\;\leq\;
c_1\: t^{-\frac{5}{6}-\varepsilon} \:+\: \frac{c_1}{t} \:+\:
\frac{C}{t}\: (\log L - \log t^{-\frac{5}{6}-\varepsilon}) \;, \]
and this has the required decay properties in $t$. \QED
In the next lemma we insert into the Fourier integral a phase
factor which oscillates infinitely fast as $\alpha \searrow 0$.
\begin{Lemma} \label{lemma41}
Let $g$ be as in Lemma~\ref{lemma40}. Then there is a constant
$c=c(g)$ such that for all $t>0$,
\begin{equation}
\left| \int_0^\infty \exp \left( i \alpha t -
\frac{i}{\sqrt{\alpha}} \right) g(\alpha)\: d\alpha \right|
\;\leq\; c\: t^{-\frac{5}{6}- \varepsilon}\;.\label{eq:4b}
\end{equation}
\end{Lemma}
{\Proof} We set
\[ \phi(\alpha) \;=\; \alpha t - \frac{1}{\sqrt{\alpha}} \;. \]
Then
\begin{equation}
\phi'(\alpha) \;=\; t + \frac{1}{2}\: \alpha^{-\frac{3}{2}}
\;,\spc \phi''(\alpha) \;=\; -\frac{3}{4}\:
\alpha^{-\frac{5}{2}}\;.\label{eq:4c}
\end{equation}
We integrate the Fourier integrals by parts,
\[ \int_0^\infty e^{i \phi(\alpha)}\: g(\alpha)\: d\alpha \;=\;
-i \int_0^\infty \left( \frac{d}{d\alpha} \:e^{i \phi(\alpha)}
\right) \frac{g}{\phi'}\: d\alpha \;=\; i \int_0^\infty e^{i
\phi(\alpha)} \left( \frac{g'}{\phi'} - \frac{g\: \phi''}{\phi'^2}
\right) d\alpha \;, \]
and obtain the bound
\begin{eqnarray*}
\lefteqn{ \left| \int_0^\infty e^{i \phi(\alpha)}\: g(\alpha)\:
d\alpha \right| \;\leq\; \int_0^\infty \left| \frac{g'}{\phi'} -
\frac{g\: \phi''}{\phi'^2} \right| \: d\alpha } \\
&=& \int_0^\infty |\phi'|^{-\frac{5}{6}-\varepsilon} \left\{
|g'|\: |\phi'|^{-\frac{1}{6}+\varepsilon} + |g|\: |\phi''|\:
|\phi'|^{-\frac{7}{6}+\varepsilon} \right\} d\alpha \;.
\end{eqnarray*}
According to~(\ref{eq:4c}), we can estimate the factor
$|\phi'|^{-\frac{5}{6}-\varepsilon}$ from above by
$t^{-\frac{5}{6}-\varepsilon}$, whereas for the factors $|\phi'|$
in the curly brackets we use the bound $|\phi'| \geq
\frac{1}{2}\:\alpha^{-\frac{3}{2}}$. Furthermore, we substitute in
the formula for $\phi''$ in~(\ref{eq:4c}) and obtain
\[ \left| \int_0^\infty e^{i \phi(\alpha)}\: g(\alpha)\: d\alpha \right|
\;\leq\;
t^{-\frac{5}{6}-\varepsilon}\: \int_0^\infty \left( c_1\: |g'|\:
\alpha^{\frac{1}{4}- \frac{3}{2}\varepsilon} \:+\: c_2\: |g|\:
\alpha^{-\frac{3}{4}- \frac{3}{2} \varepsilon} \right) d\alpha \]
with two constants $c_1$ and $c_2$. Using that $g$ is in
$L^\infty$ and $g'$ satisfies the bound~(\ref{eq:4a}), one sees
that the pole in the last integrand is integrable. \QED
The following lemma deals with the Fourier integral when we
replace the minus sign in the integrand of~(\ref{eq:4b}) by a plus
sign. Reversing this sign completely changes the long-time
asymptotics. We estimate the Fourier integral using a rigorous version of
the
``saddle point method.''

\begin{Lemma} \label{lemma42}
Let $g$ be as in Lemma~\ref{lemma40}. Then there are constants
$c=c(g)$ and $c_1=2^{-\frac{4}{3}} 3^{-\frac{1}{2}} \sqrt{\pi}$
such that for all sufficiently large $t$,
\begin{equation}
\left| \int_0^\infty \exp \left(i \alpha t +
\frac{i}{\sqrt{\alpha}} \right) g(\alpha)\: d\alpha \:-\: c_1\:
e^{i \phi_0}\: g(\alpha_0)\: t^{-\frac{5}{6}} \right| \;\leq\; c\:
t^{-\frac{5}{6}-\varepsilon}\;,\label{eq:4e}
\end{equation}
where $\alpha_0$ and $\phi_0$ are given by
\begin{equation}
\alpha_0 \;=\; (2t)^{-\frac{2}{3}} \;,\spc \phi_0 \;=\; \left(
\frac{9}{4}\: t \right)^{\frac{1}{3}}\;.    \label{eq:4f}
\end{equation}
\end{Lemma}
{\Proof}
We introduce the function $\phi$ by \[ \phi(\alpha) \;=\; \alpha t
+ \frac{1}{\sqrt{\alpha}} \;. \]
Then
\begin{equation}
\phi'(\alpha) \;=\; t - \frac{1}{2}\: \alpha^{-\frac{3}{2}}
\;,\spc \phi''(\alpha) \;=\; \frac{3}{4}\: \alpha^{-\frac{5}{2}}
\;.\label{eq:4g}
\end{equation}
One sees that $\phi(\alpha)$ has a minimum at $\alpha_0$ with
$\phi(\alpha_0)=\phi_0$ and
\begin{equation}
\phi''(\alpha_0) \;=\; 2^{-\frac{1}{3}}\:3\: t^{\frac{5}{3}} \;.
\label{eq:4h}
\end{equation}
We set
\[ \delta \;=\; t^{-\frac{5}{6}+ \varepsilon} \;. \]
For large $t$, $\delta < \alpha_0$. We split the integration range
into two regions $D_1$ and $D_2$ with
\[ D_1 \;=\; (0, \alpha_0-\delta) \cup (\alpha_0+\delta, \infty) \;,\spc
D_2 \;=\; [\alpha_0-\delta, \alpha_0+\delta] \;. \]
Let us first estimate the integral over $D_1$.  An
integration-by-parts argument similar to that in the proof of
Lemma~\ref{lemma41} gives
\[ \left| \int_{D_1} e^{i\phi}\:g\: d\alpha \right| \;\leq\;
\left| \frac{g(\alpha_0+\delta)}{\phi'(\alpha_0+\delta)} \right|
\;+\; \left| \frac{g(\alpha_0-\delta)}{\phi'(\alpha_0-\delta)}
\right| \;+\; \int_{D_1} \left| \frac{g'}{\phi'} - \frac{g\:
\phi''}{\phi'^2} \right| d\alpha\;. \]

Putting in the formulas for $\phi'$ and $\phi''$ given in (\ref{eq:4g}),
and using that $g \in L^\infty$ together with~(\ref{eq:4a}), one sees
that for suitable $c$,
\[ \left| \int_{D_1} e^{i \phi}\:g\: d\alpha \right| \;\leq\; c\:
t^{-\frac{5}{6}-\varepsilon} \;. \]
Next we show that the leading
contribution to the integral over $D_2$ is given by the saddle
point approximation. To this end, we introduce the quadratic
polynomial
\[ \phi_S(\alpha) \;=\; \phi_0 + \frac{1}{2}\: \phi''(\alpha_0)\:
(\alpha-\alpha_0)^2 \;. \]
Then the mean value theorem gives for
sufficiently large $t$,
\begin{eqnarray*}
\lefteqn{ \left| \int_{D_2} \left( e^{i \phi}\:g - e^{i \phi_S}\:
g(\alpha_0) \right) d\alpha \right| \;\leq\; \sup_{D_2}\! \left(
|g'| + |(\phi-\phi_S)'\:g|
\right) \delta^2 } \\
&\leq& \left( \frac{2C}{\alpha_0} + \frac{1}{2}\: \sup_{D_2}
|\phi'''|\: \delta^2\: \|g\|_\infty \right) \delta^2 \;\leq\;
\left( c_2\: t^{-1+2\varepsilon} + c_3\: t^{-1+4 \varepsilon}
\right) \;\leq\; c\: t^{-\frac{5}{6}-\varepsilon} \;,
\end{eqnarray*}
where in the last step we used that $5 \varepsilon < \frac{1}{6}$.
Finally, we compute the contribution of the saddle point
approximation.
\[ \int_{D_2} e^{i \phi_S}\: g(\alpha_0)\: d\alpha \;=\; e^{i \phi_0}\:
g(\alpha_0) \int_{-\delta}^\delta e^{\frac{i}{2}\:
\phi''(\alpha_0)\: \alpha^2}\: d\alpha \;. \]
Introducing the new
variable $s=\frac{1}{2}\: \phi''(\alpha_0)\:\alpha^2$ gives
\[ \int_{D_2} e^{i \phi_S}\: g(\alpha_0)\: d\alpha \;=\;
e^{i \phi_0}\: g(\alpha_0)\: \sqrt{\frac{2}{\phi''(\alpha_0)}}
\int_0^L \frac{e^{is}}{\sqrt{s}}\:ds \]
with
\begin{equation}
L \;=\; \frac{1}{2}\: \phi''(\alpha_0)\: \delta^2 \;=\;
2^{-\frac{4}{3}}\:3\: t^{2 \varepsilon}\;.    \label{eq:4j}
\end{equation}
Using~(\ref{eq:4h}), we conclude that
\[ \int_{D_2} e^{i \phi_S}\: g(\alpha_0)\: d\alpha \;=\; e^{i \phi_0}\:
g(\alpha_0)\: t^{-\frac{5}{6}}\: (c_1 + R(t)) \]
with
\begin{eqnarray*}
c_1 &=& 2^{\frac{1}{3}}\: 3^{-\frac{1}{2}} \int_0^\infty
\frac{e^{is}}{\sqrt{s}}\:ds \:=\:2^{\frac{4}{3}}\:
3^{-\frac{1}{2}}\sqrt{\pi}  \;\neq\; 0 \\
R(t) &=& -2^{\frac{1}{3}}\: 3^{-\frac{1}{2}} \int_L^\infty
\frac{e^{is}}{\sqrt{s}}\:ds \;.
\end{eqnarray*}
The error term $R(t)$ can be integrated by parts,
\[ \left| \int_L^\infty \frac{e^{is}}{\sqrt{s}}\:ds \right| \;\leq\;
\frac{1}{\sqrt{L}} + \frac{1}{2} \int_L^\infty
s^{-\frac{3}{2}}\:ds \;=\; \frac{5}{4}\: \frac{1}{\sqrt{L}}\;, \]
and this shows according to~(\ref{eq:4j}) that $R(t)$ decays in
$t$ at the desired rate, $|R(t)| \leq c\: t^{-\varepsilon}$.
\QED

\section{The Planar Equation}
\setcounter{equation}{0}
Let us transform the radial Dirac equation~(\ref{eq:a}) into an
equation for a real $2$-spinor as follows. We first unitarily
transform the spinor $X$ according to
\begin{equation}
X \rightarrow \tilde{X}=UX \spc{\mbox{with}}\spc U \;=\; \exp
\!\left( i \:\frac{\beta}{2}\:\sigma^3 \right) \;,\;\;\;\;\; \beta
\;=\; \arctan \frac{\lambda}{mr} \;. \label{eq:19a}
\end{equation}
Then $\tilde{X}$ satisfies the equation
\begin{equation}
\frac{d}{du} \tilde{X} \;=\; i \left( \begin{array}{cc} -a & b \\
-b & a
\end{array} \right) \tilde{X}
\label{eq:2yy}
\end{equation}
with
\begin{eqnarray}
a(u) & = & \Omega(u) \:+\: \frac{\lambda m}{m^2 r^2 + \lambda^2}
\:\frac{\Delta}{2 (r^2+a^2)} \label{eq:2xx} \\
b(u) &=& \frac{\sqrt{\Delta}}{r^2+a^2}\:\sqrt{m^2 r^2 + \lambda^2}
\;. \label{eq:2w}
\end{eqnarray}
Notice that the transformation $U$ is regular for all $u \in \R$,
and that the second summand in~(\ref{eq:2xx}) has nice decay
properties for $u \rightarrow \pm \infty$. Next we employ the
ansatz
\begin{equation}
\tilde{X} \;=\; \left( \begin{array}{c} \Gamma(a) \:\psi^+ \:-\: i \psi^- \\
-\psi^+ \:-\: i \:\Gamma(a)\:\psi^- \end{array} \right)
\;\;\;,\spc \psi \;=\; \frac{1}{2} \left( \begin{array}{c}
\Gamma(a) \:\tilde{X}^+
\:-\: \tilde{X}^- \\
i \tilde{X}^+ \:+\: i \:\Gamma(a)\:\tilde{X}^- \end{array}
\right) \label{eq:2phi}
\end{equation}
with a complex $2$-spinor $\psi$. Then $\psi$ satisfies the
equation
\begin{equation}
\frac{d}{du} \psi \;=\; \left( \begin{array}{cc} 0 & -g \\ f & 0
\end{array}
\right) \psi \;, \label{eq:2a}
\end{equation}
with
\begin{equation}
f \;=\; |a|+b \;\;\;,\spc g \;=\; |a|-b \;. \label{eq:23a}
\end{equation}
The coefficients in~(\ref{eq:2a}) are all real, and so we can
study the real and imaginary parts of $\psi$ separately. Thus we
assume in what follows that $\psi$ is real and then
call~(\ref{eq:2a}) the {\em{planar equation}}.

We bring the planar equation into a form more appropriate for our
estimates. For given $u_0$ we introduce the new variable
\begin{equation}
x(u) \;=\; 2 \int_{u_0}^u \sqrt{|fg|(\tau)}\: d\tau
\label{eq:2var}
\end{equation}
and set
\begin{equation}
h \;=\; \frac{1}{2}\: \log \left| \frac{g}{f} \right| \;.
\label{eq:24a}
\end{equation}
In the case $g>0$, (\ref{eq:2a}) becomes
\[ \psi' \;\equiv\; \frac{d}{dx} \:\psi \;=\; \frac{1}{2}\: \left(
\begin{array}{cc}
0 & -e^h \\ e^{-h} & 0 \end{array} \right) \psi \;. \] Employing
the ansatz
\begin{equation}
\psi \;=\; e^{-\frac{L}{2}}\: \left( \begin{array}{cc}
e^{\frac{h}{2}}\: \cos \frac{x+\vartheta}{2} \\ e^{-\frac{h}{2}}\:
\sin \frac{x+\vartheta}{2} \end{array} \right) \label{eq:25a}
\end{equation}
with real functions $L(x)$ and $\phi(x)$ gives the equation
\begin{equation}
-L'\: a_1 \:+\: h'\: b \:+\: (1+\vartheta')\: a_2 \;=\; a_2
\label{eq:2z}
\end{equation}
with
\[ a_1 \;=\; \left( \begin{array}{c} \cos \frac{x+\vartheta}{2} \\[.2em]
\sin
\frac{x+\vartheta}{2} \end{array} \right) \;,\spc
a_2 \;=\; \left( \begin{array}{c} -\sin \frac{x+\vartheta}{2} \\[.2em]
\cos \frac{x+\vartheta}{2} \end{array} \right) \;,\spc
b \;=\; \left( \begin{array}{c} \cos \frac{x+\vartheta}{2} \\[.2em]
-\sin \frac{x+\vartheta}{2}
\end{array} \right) \;. \]
Elementary trigonometry shows that
\[ b \;=\; \cos (x+\vartheta)\: a_1 \:-\: \sin (x+\vartheta)\: a_2 \;. \]
Hence the planar equation takes the form
\begin{equation}
\vartheta' \;=\; h'\: \sin (x+\vartheta) \;\;\;,\spc L' \;=\; h'\:
\cos (x+\vartheta) \;. \label{eq:2y}
\end{equation}
In the case $g<0$, the ansatz
\begin{equation}
\psi \;=\; e^{-\frac{L}{2}}\: \left( \begin{array}{cc}
e^{\frac{h}{2}}\: \cosh \frac{x+\vartheta}{2} \\
e^{-\frac{h}{2}}\: \sinh \frac{x+\vartheta}{2} \end{array} \right)
\label{eq:an2}
\end{equation}
gives similarly the equations
\begin{equation}
\vartheta' \;=\; h'\: \sinh (x+\vartheta) \;\;\;,\spc L' \;=\;
h'\: \cosh (x+\vartheta) \;. \label{eq:2x}
\end{equation}

We can now give the strategy for the proof of Theorem~\ref{thm1}{\bf{(i)}}. 
First, in the next section, we will obtain estimates which will enable 
us to control the function $h'$ which appears in the planar 
equations~(\ref{eq:2y}) and~(\ref{eq:2x}). Then we will carefully 
analyze the solutions $(\vartheta, L)$ of these planar equations, 
and this will allow us to study the time-dependence of the 
propagator~(\ref{eq:1b}). For the analysis of the planar equations, it is
necessary to  consider both cases $\omega>m$ and $\omega<m$ separately; this
will be  done in Sections~\ref{sec6} and~\ref{sec7}, respectively.

\section{Uniform Bounds for the Potentials} \label{sec5}
\setcounter{equation}{0}
In this section, we shall derive estimates for the function $h(x)$
(as introduced in~(\ref{eq:24a}) with $x$ according
to~(\ref{eq:2var})) as well as for its partial derivatives with
respect to $x$ and $\omega$. The usefulness of our estimates lies
in the fact that they are uniform in $\omega$ for $\omega$ in a
small neighborhood of $m$, $\omega \in (m-\delta, m+\delta)$. The
main technical difficulty is that $x$ is defined via an integral
transformation~(\ref{eq:2var}), and thus $h(x)$ depends on
$\omega$ in a nonlocal way. On the other hand, our estimates also
show the advantage of working with the variable $x$. Namely, by
introducing $x$, the $\omega$-dependence of $h$ becomes small in
the critical regions near infinity and near the poles of $h$, in the
sense that $(\omega-m) \partial_\omega h(x)$ has bounded total
variation in $x$, uniformly in $\omega$. This will be essential
for getting control of the $\omega$-dependence of the solutions to
the planar equation (see Lemmas~\ref{lemma62} and~\ref{lemma75}). Since the
technical details of the proofs of Lemmas~\ref{lemma51}, \ref{lemma52},
and~\ref{lemma52a} will not be needed later on, the reader may consider
skipping these proofs in a first reading.

In what follows, we often denote derivatives by a lower
index, e.g.\ $h_\omega \equiv \partial_\omega h$.
Furthermore, we denote constants which are
independent  of $\omega$ by $c$; the value of $c$ may change
throughout our calculations.  For clarity, we sometimes add a
subscript to $c$ to mean a fixed constant. According to
their definition~(\ref{eq:23a})  and~(\ref{eq:2xx}),(\ref{eq:2w}),
the functions $f$ and $g$ have for large $u$  the expansion
\begin{eqnarray*}
f &=& (\omega+m) \:-\: \frac{mM - eQ}{u} \:+\:
{\cal{O}}\!\left(\frac{1}{u^2}\right) \\
g &=& (\omega-m) \:+\: \frac{mM + eQ}{u} \:+\:
{\cal{O}}\!\left(\frac{1}{u^2}\right) \;.
\end{eqnarray*}
Our notation ${\cal{O}}(u^{-n})$ implies that the error terms
depend smoothly on $\omega$, and that their $u$-derivatives have
the natural scaling behavior, i.e.
\[ \partial_\omega {\cal{O}}(u^{-n}) = {\cal{O}}(u^{-n})
\spc {\mbox{and}} \spc
\partial_u {\cal{O}}(u^{-n}) = {\cal{O}}(u^{-n-1}) \;. \]
Our assumption~(\ref{eq:cc}) ensures that for large $u$, $g$ is
monotone decreasing, whereas $f$ is increasing.

We begin with the case $\omega>m$. In this parameter range, we fix
$u_0$ independent of $\omega$. By choosing $u_0$ sufficiently
large, we can arrange that the following estimates hold.

\begin{Lemma} \label{lemma51}
There are constants $c, \delta>0$ such that for all $\omega \in
(m, m+\delta)$ and $x>0$,
\begin{eqnarray}
0 \;<\; -h'(x) &\leq& \frac{c}{1+x} \label{eq:5a} \\
|h''(x)| &\leq& \frac{c}{(1+x)^2} \label{eq:5b} \\
\int_0^\infty |h'_\omega(x)|\:dx &\leq& \frac{c}{\omega-m}\;.
\label{eq:5c}
\end{eqnarray}
\end{Lemma}
{\Proof} We set $\varepsilon = \omega^2-m^2$ and introduce the
function
\[ \rho = 2 \sqrt{fg} \;. \]
Then $h$ and $\rho$ have the asymptotic expansions
\begin{eqnarray}
h(u) &=& \frac{1}{2}\: \log \left(
\frac{\varepsilon}{(\omega+m)^2} + \frac{\alpha}{u} +
{\cal{O}}\!\left( \frac{1}{u^2} \right) \right)
\label{eq:5A} \\
\rho(u) &=& 2\: \sqrt{\varepsilon + \frac{\beta}{u}
 + {\cal{O}}\!\left( \frac{1}{u^2} \right)} \label{eq:5B}
\end{eqnarray}
with positive constants $\alpha$ and $\beta$, which depend
smoothly on $\omega$ and are bounded away from zero as
$\varepsilon \rightarrow 0$. Our first step is to bound the
function $x(u)$, (\ref{eq:2var}), as well as its inverse $u(x)$.
According to~(\ref{eq:5B}), there are (possibly after increasing
$u_0$), constants $a_1, a_2>0$ such that
\begin{equation}
2 \:\sqrt{\varepsilon + \frac{a_1}{u}} \;\leq\; \rho(u) \;\leq\; 2
\:\sqrt{\varepsilon + \frac{a_2}{u}} \label{eq:5C}
\end{equation}
for all $\omega \in (m, m+\delta)$ and $u>u_0$. We introduce the
functions $\underline{x}$ and $\overline{x}$ by
\[ \underline{x} \;=\; 2 \sqrt{u\:(a_1+\varepsilon u)} - b_1\;,\spc
\overline{x} \;=\; 4 \sqrt{u\:(a_2+\varepsilon u)} - b_2 \;, \]
where the constants $b_1$ and $b_2$ are chosen such that
$\underline{x}(u_0)=0=\overline{x}(u_0)$,
\begin{equation}
b_1 \;=\; 2 \sqrt{u_0\:(a_1+\varepsilon u_0)} \;,\spc b_2 \;=\; 4
\sqrt{u_0\:(a_2+\varepsilon u_0)}\;.    \label{eq:5Y}
\end{equation}
Then
\begin{eqnarray*}
\underline{x}'(u) &=& \frac{a_1+2 \varepsilon
u}{\sqrt{u(a_1+\varepsilon u)}}
\;\leq\; 2 \sqrt{\frac{a_1+\varepsilon u}{u}} \;\leq\; \rho(u) \\
\overline{x}'(u) &=& \frac{2(a_2+2 \varepsilon
u)}{\sqrt{u(a_2+\varepsilon u)}} \;\geq\; 2
\sqrt{\frac{a_2+\varepsilon u}{u}} \;\geq\; \rho(u) \;,
\end{eqnarray*}
and integration yields that $\underline{x}$ and $\overline{x}$ are
bounds for $x$,
\begin{equation}
\underline{x}(u) \;\leq\; x(u) \;\leq\; \overline{x}(u) \spc
{\mbox{for all $u \geq u_0$.}} \label{eq:5D}
\end{equation}
The functions $\underline{x}$ and $\overline{x}$ are strictly
monotone and thus invertible. Their inverses are computed as
follows,
\begin{eqnarray*}
\underline{x}^{-1}(x) &=& \frac{1}{2 \varepsilon} \left(
\sqrt{a_1^2 + \varepsilon\: (x+b_1)^2} - a_1 \right) \\
&=& \frac{1}{2 \varepsilon}\: \frac{\varepsilon\: (x+b_1)^2}
{\sqrt{a_1^2 + \varepsilon\: (x+b_1)^2} + a_1} \;\leq\;
\frac{1}{2}\: \frac{(x+b_1)^2}{\sqrt{\varepsilon}\:(x+b_1) + a_1} \\
\overline{x}^{-1}(x) &=& \frac{1}{4 \varepsilon} \left(
\sqrt{4 a_2^2 + \varepsilon\: (x+b_2)^2} - 2a_2 \right) \\
&=& \frac{1}{4 \varepsilon}\: \frac{\varepsilon\: (x+b_2)^2}
{\sqrt{4 a_2^2 + \varepsilon\: (x+b_2)^2} + 2a_2} \;\geq\;
\frac{1}{4}\: \frac{(x+b_2)^2}{\sqrt{\varepsilon}\:(x+b_2) + 4
a_2} \;,
\end{eqnarray*}
where in the last step we applied the inequality $\sqrt{a+b} \leq
\sqrt{a} +\sqrt{b}$ ($a,b>0$). The inequalities~(\ref{eq:5D})
yield for the inverses that $\underline{x}^{-1}(x) \geq u(x) \geq
\overline{x}^{-1}(x)$. Thus the functions $\underline{u}$ and
$\overline{u}$ defined by
\begin{equation}
\underline{u}(x) \;=\; \frac{1}{4}\:
\frac{(x+b_2)^2}{\sqrt{\varepsilon}\:(x+b_2) + 4 a_2} \;,\spc
\overline{u}(x) \;=\; \frac{1}{2}\:
\frac{(x+b_1)^2}{\sqrt{\varepsilon}\:(x+b_1) + a_1} \label{eq:5E}
\end{equation}
are bounds for $u(x)$,
\begin{equation}
\underline{u}(x) \;\leq\; u(x) \;\leq\; \overline{u}(x) \spc
{\mbox{for all $x \geq 0$.}} \label{eq:5F}
\end{equation}

\begin{figure}[tb]
    \epsfxsize=16cm
    \centerline{\epsfbox{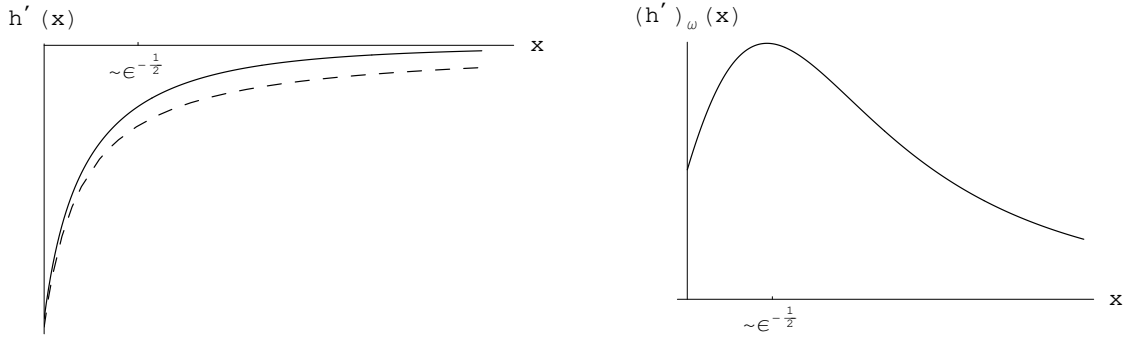}}
    \caption{Typical plots for $h'(x)$ and $h'_\omega(x)$
in the case $\omega>m$.}
    \label{fig2}
\end{figure}

Let us derive~(\ref{eq:5a}). Since $f$ and $g$ are monotone
increasing and decreasing, respectively, $h(u)$ is decreasing and
thus $h'(x)<0$. Furthermore,
\begin{eqnarray}
|h'(x)| &=& \frac{1}{\rho}\: |h'(u)| \;=\; \frac{1}{2 \rho} \left|
\frac{g'(u)}{g} - \frac{f'(u)}{f} \right| \nonumber \\
&=& \frac{2}{\rho^3}\: |g'(u)\: f - f'(u)\: g| \;\leq\;
\frac{c_1}{u^2\: \rho^3} \;. \label{eq:59b}
\end{eqnarray}
We employ~(\ref{eq:5C}), (\ref{eq:5F}), and~(\ref{eq:5E}) to
obtain
\begin{eqnarray*}
|h'(x)| &\leq& \frac{c_1}{8}\: u^{-\frac{1}{2}}\: (a_1 +
\varepsilon u)^{-\frac{3}{2}} \;\leq\; \frac{c_1}{8}\:
\underline{u}^{-\frac{1}{2}}\:
(a_1 + \varepsilon \underline{u})^{-\frac{3}{2}} \\
&=& 2c_1 \left( \frac{\sqrt{\varepsilon} (x+b_2)+4a_2}{(x+b_2)^2}
\right)^{\frac{1}{2}} \left( \frac{\sqrt{\varepsilon}\:
(x+b_2)+4a_2}{\varepsilon (x+b_2)^2 +4 a_1\: \sqrt{\varepsilon}
(x+b_2)+16 a_1 a_2}
\right)^{\frac{3}{2}} \\
&\leq& \frac{2c_1}{x+b_2} \left[
\frac{(\sqrt{\varepsilon}\:(x+b_2) + 4 a_2)^2} {(\varepsilon
(x+b_2)^2 + 16 a_1 a_2)^\frac{3}{2}} \right] \;.
\end{eqnarray*}
The square bracket is bounded uniformly in $\varepsilon$ and $x$,
proving~(\ref{eq:5a}). The second derivative of $h$ is computed to
be
\begin{eqnarray*}
h''(x) &=& \frac{1}{\rho} \:\frac{d}{du} \left(
\frac{2}{\rho^3}\: (g'(u)\: f - f'(u)\: g) \right) \\
&=& -\frac{6}{\rho^5} \left( \rho'(u)\:(g'(u)\: f - f'(u)\: g)
\right) + \frac{2}{\rho^4} \: \frac{d}{du} (g'(u)\: f - f'(u)\:
g)\;.
\end{eqnarray*}
Thus
\begin{equation}
|h''(x)| \;\leq\; \frac{c_2}{u^4\: \rho^6} + \frac{c_2}{u^3\:
\rho^4}\;, \label{eq:51b}
\end{equation}
and~(\ref{eq:5C}) shows that
\[ |h''(x)| \;\leq\; 8c_2\: u^{-1}\:(a_1+\varepsilon u)^{-3}
+ c_2\:u^{-1}\: (a_1 +\varepsilon u)^{-2} \;\leq\; c_3\:
\underline{u}^{-1} \: (a_1 + \varepsilon \underline{u})^{-2}\;. \]
We substitute in~(\ref{eq:5E}),
\[ |h''(x)| \;\leq\; \frac{16 c_3}{(x_1+b_1)^2} \left[
\frac{ (\sqrt{\varepsilon} (x+b_2) + 4 a_2)^3} {(\varepsilon
\:(x+b_2)^2 + 16 a_1 a_2)^2} \right] \;. \] The square bracket is
again bounded uniformly in $\varepsilon$ and $x$, and this
gives~(\ref{eq:5b}). We finally estimate $h'_\omega(x)$.  Since
the relation between $\omega$ and $\varepsilon$ is one-to-one and
smooth, we can just as well consider the $\varepsilon$-derivative
$h'_\varepsilon(x)$.  Since $h(x)$ is not given in closed form, we
need to compute $h_\varepsilon(x)$ via the formula
\begin{equation}
\varepsilon \:h_\varepsilon(x) \;=\; \varepsilon \:h_\varepsilon(s) +
\varepsilon \: h'(s)\:
s_\varepsilon(x)\;, \label{eq:5G}
\end{equation}
where $s=s(u)$ is a suitable variable.  Clearly,
$h_\varepsilon(x)$ is independent of how $s$ is chosen.  However,
if we take for $s$ too simple a function (e.g.\ $s=u$), then
it turns out that $h_\varepsilon(s)$ will develop singularities in the
limit $\varepsilon \rightarrow 0$, which are compensated in~(\ref{eq:5G})
by corresponding singular contributions to the second summand, making
the analysis very delicate.  To bypass these difficulties, it is
convenient to choose for $s(u)$ a function with a similar
qualitative behavior as $x(u)$; this will make it possible to
estimate the two summands in~(\ref{eq:5G}) separately. We set
$s=[{\mbox{$\overline{u}$-inverse}}+b_1]$, so that 
$u=\overline{u}(s(u)-b_1)$; i.e.
\[ h(s) \;=\; h(u=\overline{u}(s-b_1)) \]
with $\overline{u}$ according to~(\ref{eq:5E}) and $s \in [b_1,
\infty)$. Then the expansion~(\ref{eq:5A}) becomes
\[ h(s) \;=\; \frac{1}{2}\: \log \left( \varepsilon \:\lambda_1 +
\frac{2 \lambda_2 \sqrt{\varepsilon}}{s} + \frac{\lambda_3}{s^2} +
\sqrt{\varepsilon}\:{\cal{O}}(s^{-2}) + {\cal{O}}(s^{-3}) \right) \]
with positive constants $\lambda_i$ which depend smoothly on
$\sqrt{\varepsilon}$ and are uniformly bounded away from zero.
Differentiating with respect to $\varepsilon$ gives
\[ \varepsilon\: h_\varepsilon(s) \;=\; \frac{1}{2}\:
\frac{\lambda_1 \:\varepsilon s^2 + \lambda_2 \:\sqrt{\varepsilon}
s +\sqrt{\varepsilon}\: {\cal{O}}(s^0)} {\lambda_1 \:\varepsilon
s^2 + 2 \lambda_2 \:\sqrt{\varepsilon} s + \lambda_3
+\sqrt{\varepsilon}\:{\cal{O}}(s^0)+{\cal{O}}(s^{-1})} \;. \]
We want to show that this function has bounded total variation. To
this end, we differentiate with respect to $s$ and obtain
\[ \varepsilon\: h'_\varepsilon(s) \;=\;
\frac{ c_1 \:\varepsilon^{\frac{3}{2}} s^2 + c_2 \:\varepsilon s +
c_3\: \sqrt{\varepsilon} + \varepsilon^{\frac{3}{2}}\:
{\cal{O}}(s) + \varepsilon\: {\cal{O}}(s^0) + \sqrt{\varepsilon}\:
{\cal{O}}(s^{-1})} {2 \:(\lambda_1\: \varepsilon s^2 + 2 \lambda_2
\:\sqrt{\varepsilon} s + \lambda_3 +
\sqrt{\varepsilon}\:{\cal{O}}(s^0)+{\cal{O}}(s^{-1}))^2} \;. \]
Hence by choosing $\delta$ small enough and $u_0$ (and thus $b_2$)
large enough, we can arrange that
\[ \varepsilon \:|h'_\varepsilon| \;\leq\;
c\: \frac{\varepsilon^{\frac{3}{2}} s^2 + \varepsilon s +
\sqrt{\varepsilon}} {(\lambda_1 \varepsilon s^2 + \lambda_3)^2} \]
for all $\varepsilon$ and all $s \geq b_1$. The $L^1$ norm of the rhs
is bounded uniformly in $\omega$. Namely, setting $t=\sqrt{\lambda_1 
\varepsilon}\: s$ shows that for $n=0,1,2$,
\[ \int_{b_2}^\infty \frac{(\sqrt{\varepsilon} s)^n}{(\lambda_1 \varepsilon
s^2 + \lambda_3)^2} \:\sqrt{\varepsilon}\:ds \;\leq\;
\lambda_1^{-\frac{n+1}{2}} \int_0^\infty \frac{t^n}{(t^2+
\lambda_3)^2} \: dt \; \] and the last integral is finite,
independent of $\varepsilon$. It remains to estimate the total
variation of the second summand in~(\ref{eq:5G}). More precisely,
in order to finish the proof of~(\ref{eq:5c}), we shall show that
\begin{equation}
\int_0^\infty \left| \frac{d}{dx} (h'(s)\;\varepsilon
\:s_\varepsilon(x)) \right| dx \;<\; c\;. \label{eq:5H}
\end{equation}
We first derive sufficient conditions for~(\ref{eq:5H}). The
relations
\[ 0 \;=\; \partial_\varepsilon s(x(s)) \;=\; s_\varepsilon(x) + s'(x)\:
x_\varepsilon(s) \,\spc h'(x) \;=\; h'(s)\: s'(x) \]
yield that
\[ h'(s)\: s_\varepsilon(x) \;=\; - h'(x)\: x_\varepsilon(s) \;. \]
Differentiating with respect to $x$, one sees that it suffices to
bound the $L^1$ norms of the expressions
\[ h'(x)\: \frac{\varepsilon \:x'_\varepsilon(s)}{x'(s)}
\spc{\mbox{and}}\spc
h''(x)\: \varepsilon \:x_\varepsilon(s) \] uniformly in
$\varepsilon$. Substituting in the bounds~(\ref{eq:5a})
and~(\ref{eq:5b}), we conclude that the following inequalities
imply that~(\ref{eq:5H}) holds,
\begin{eqnarray}
\frac{1}{c} \;\leq\; x'(s) &\leq& c \label{eq:5I} \\
\int_0^\infty \frac{\varepsilon\: |x'_\varepsilon(s)|}{1+x}\:dx
&\leq& c \label{eq:5K} \\
\int_0^\infty \frac{\varepsilon\: |x_\varepsilon(s)|}{(1+x)^2}\:dx
&\leq& c \;. \label{eq:5J}
\end{eqnarray}
We begin the proof of~(\ref{eq:5I})--(\ref{eq:5J}) by computing
$x'(s)$,
\begin{equation}
x'(s) \;=\; \frac{d}{ds} \int_{u_0}^{\overline{u}(s-b_1)}
\rho(v)\:dv \;=\; \rho(\overline{u}(s-b_1))\: \overline{u}'(s-b_1)
\;. \label{eq:5Z}
\end{equation}
A short calculation using~(\ref{eq:5E}) and~(\ref{eq:5B}) gives
\begin{eqnarray*}
\overline{u}'(s-b_1) &=& \frac{s(\sqrt{\varepsilon} s + 2 a_1)}
{2 (\sqrt{\varepsilon} s + a_1)^2} \;>\; 0 \\
\rho^2(\overline{u}(s-b_1)) &=& \frac{1}{s^2}\: (4 \varepsilon s^2
+ 8 \beta \sqrt{\varepsilon} s + 8 \beta a_1)
+\sqrt{\varepsilon}\: {\cal{O}}(s^{-2}) + {\cal{O}}(s^{-3})
\end{eqnarray*}
and thus
\begin{equation}
x'(s)^2 \;=\; \frac{ (\sqrt{\varepsilon} s)^4 + a_3
\:(\sqrt{\varepsilon} s)^3 + \cdots + a_0 \:+\:
\varepsilon^{\frac{3}{2}}\:{\cal{O}}(s^{2}) + \varepsilon\:
{\cal{O}}(s) + \sqrt{\varepsilon}\:{\cal{O}}(s^0) +
{\cal{O}}(s^{-1}) }
{ (\sqrt{\varepsilon} s)^4 + b_3 \:(\sqrt{\varepsilon} s)^3 +
\cdots + b_0} \label{eq:5L}
\end{equation}
with coefficients $a_j, b_j \geq 0$ and $a_0, b_0 >0$. Possibly
after increasing $u_0$ and decreasing $\delta$, we can neglect the
error terms. The fraction in~(\ref{eq:5L}) is clearly uniformly
bounded from above and below. This proves~(\ref{eq:5I}).
Integrating~(\ref{eq:5I}), we obtain that the ratio
$(1+x)/s$ is uniformly bounded from above and below, and thus
in~(\ref{eq:5K}) and (\ref{eq:5J}) we may replace the factors $(1+x)$
by $s$. Using~(\ref{eq:5I}) we may furthermore replace
the integral over $x \in (0, \infty)$ by the integral
over $s \in (b_1, \infty)$. Next we differentiate~(\ref{eq:5L})
with respect to $\varepsilon$. A short computation shows that
\[ \varepsilon \:x'_\varepsilon(s)\: x'(s) \;=\;
\frac{a_7\:(\sqrt{\varepsilon} s)^7 + \cdots + a_1
\:(\sqrt{\varepsilon} s)} {(\sqrt{\varepsilon} s)^8 +
b_7\:(\sqrt{\varepsilon} s)^7 + \cdots + b_0} \left( 1 +
{\cal{O}}(s^{-1}) \right) , \] where the coefficients $b_j$ are
non-negative and $b_0 >0$ (but the $a_j$ might be zero or
negative). Using the bounds~(\ref{eq:5I}), one sees that
$\varepsilon x'_\varepsilon(s)$ can be estimated by
\begin{equation}
\varepsilon\: |x'_\varepsilon(s)| \;\leq\; c \:\frac{
\sqrt{\varepsilon} s \: ((\sqrt{\varepsilon}
s)^6+1)}{(\sqrt{\varepsilon} s)^8 + 1} \;. \label{eq:5M}
\end{equation}
A scaling argument shows that
\begin{equation}
\int_{b_1}^\infty \frac{\varepsilon\: |x'_\varepsilon(s)|}{s}\: ds
\;\leq\; c \int_{b_1}^\infty \frac{(\sqrt{\varepsilon} s)^6+1}
{(\sqrt{\varepsilon} s)^8 + 1}\: \sqrt{\varepsilon} \:ds \;\leq\;
c \int_0^\infty \frac{t^6+1}{t^8+1} \;<\; \infty \;, \label{eq:5N}
\end{equation}
proving~(\ref{eq:5K}). To derive~(\ref{eq:5J}), we use that
\[ \varepsilon \:|x_\varepsilon(s)| \;\leq\; \int_{b_1}^s \varepsilon\:
|x'_\varepsilon(t)|\:dt \:+\: \varepsilon \:|x_\varepsilon(b_1)| \]
and obtain
\begin{eqnarray*}
\int_{b_1}^\infty \frac{\varepsilon\: |x_\varepsilon(s)|}{s^2}\:ds
&\leq& \int_{b_1}^\infty \frac{1}{s^2} \left( \int_{b_1}^s
\varepsilon\:
|x'_\varepsilon(t)|\: dt \:+\: \varepsilon \:|x_\varepsilon(b_1)| \right) ds
\\
&=& - \int_{b_1}^\infty \frac{d}{ds} \left(\frac{1}{s}\right)
\left( \int_{b_1}^s \varepsilon\: |x'_\varepsilon(t)|\: dt \:+\:
\varepsilon \:|x_\varepsilon(b_1)| \right) ds
\end{eqnarray*}
According to~(\ref{eq:5M}), the inner integral diverges at most
logarithmically as $s \rightarrow \infty$. Therefore, integrating
by parts gives no boundary terms at infinity,
\[ \int_{b_1}^\infty \frac{\varepsilon\: |x_\varepsilon(s)|}{s^2}\:ds
\;\leq\; \int_{b_1}^\infty
\frac{\varepsilon\:|x'_\varepsilon(s)|}{s}\: ds \:+\:
\frac{\varepsilon\: |x_\varepsilon(b_1)|}{b_1}\;. \]
The integral
on the right was estimated in~(\ref{eq:5N}). The last
summand is computed to be
\[ \varepsilon\:\frac{|x_\varepsilon(b_1)|}{b_1} \;=\;
\varepsilon\: \frac{|x'(b_1)|}{b_1}\: \frac{\partial b_1}{\partial 
\varepsilon} \;, \]
and this is bounded uniformly in $\varepsilon$ in view
of~(\ref{eq:5Z}) and the fact that $b_1$ is smooth in
$\sqrt{\varepsilon}$ and bounded away from zero, (\ref{eq:5Y}).
This completes the proof of Lemma~\ref{lemma51}.
\QED
The above estimates are illustrated in Figure~\ref{fig2},
where $h'$ and $h'_\omega$ are plotted in a typical example.
The dashed curve describes the asymptotics near $x=0$; it is the
graph of $h'(x)$, where for $x(u)$ one uses the approximate
formula $x \approx 4 \sqrt{\beta} (\sqrt{u} - \sqrt{u_0})$,
obtained by setting $\varepsilon$ in~(\ref{eq:5B}) equal to zero,
dropping the error term and integrating, (\ref{eq:2var}).

In the case $\omega<m$, we fix $\umin$ independent of $\omega \in
(m-\delta, m)$.  By choosing $\umin$ large and $\delta$
sufficiently small, we can arrange that the function $g$ has
exactly one zero on the half line $(\umin, \infty)$. We set $u_0$
equal to this zero,
\[ g(u_0) \;=\; 0\;. \]
Clearly, $u_0$ depends on $\omega$.  The variable $x(u)$,
(\ref{eq:2var}), is positive for $u>u_0$ and negative on the
interval $(\umin, u_0)$. We set $\xmin = x(\umin)$. The following
lemma is the analogue of Lemma~\ref{lemma51} for $\omega<m$. The
method of proof is also similar, but the pole of $h$ at $x=0$
makes the situation a bit more complicated.

\begin{Lemma} \label{lemma52}
There are constants $c, \delta>0$ such that for all $\omega \in
(m-\delta, m)$ and $x>\xmin$,
\begin{eqnarray}
0 \;\leq\; \Gamma(x)\: h'(x) &\leq&
\frac{c}{x-\xmin+1} \:+\: \frac{c}{|x|} \label{eq:56} \\
|h''(x)| &\leq& \frac{c}{(x-\xmin+1)^2} \:+\: \frac{c}{|x|^2} \label{eq:57}
\\
\int_{\sxmin}^\infty |h'_\omega(x)|\: dx &\leq&
\frac{c}{m-\omega}\;. \label{eq:58}
\end{eqnarray}
Furthermore, for every $x_1>0$ the constants $c, \delta>0$ can be
chosen such that for all $\omega \in (m-\delta, m)$ and $x \in
[-x_1, x_1]$,
\begin{equation}
\left| h(x) - \frac{1}{3}\: \log |\varepsilon^2 x| \right| \;\leq\;
c\;. \label{eq:5nasy}
\end{equation}
\end{Lemma}
{\Proof} We now set $\varepsilon = m^2-\omega^2$ and $\rho = 2
\sqrt{|fg|}$. Then
\[ \rho^2(u) \;=\; 4 \left| \varepsilon - \frac{\beta}{u} +
{\cal{O}}(u^{-2})
\right| \]
with $\beta>0$. Since $\rho(u_0)=0$,
\begin{equation}
u_0 \;=\; \frac{\beta}{\varepsilon} \:(1 +
{\cal{O}}(\varepsilon))\;, \label{eq:509}
\end{equation}
and furthermore, the functions $h$ and $\rho$ have the expansions
\begin{eqnarray}
h(u) &=& \frac{1}{2}\: \log \left(
\frac{\varepsilon}{(\omega+m)^2}\:
\frac{|u-u_0|}{u} \left(1+{\cal{O}}(u^{-1}) \right) \right) \label{eq:510}
\\
\rho(u) &=& 2 \:\sqrt{\varepsilon\: \frac{|u-u_0|}{u}
\left(1+{\cal{O}}(u^{-1}) \right)}\;. \label{eq:511}
\end{eqnarray}
Since global bounds for $x(u)$ and $u(x)$ would be more difficult to 
obtain than those in Lemma~\ref{lemma51}, we here construct
the bounds piecewise.  We set $\Delta u = u_0/2$.  By decreasing
$\delta$, we can arrange that $u_0/4>u_{\mbox{\scriptsize{min}}}$,
and furthermore we can also make the error terms in~(\ref{eq:510})
and~(\ref{eq:511}) as small as we like.  Thus we may assume that
\begin{equation}
\frac{3}{2}\: \sqrt{\frac{\varepsilon}{u_0}}\: \sqrt{|u-u_0|}
\;\leq\; \rho(u) \;\leq\; 3\: \sqrt{\frac{\varepsilon}{u_0}}\:
\sqrt{|u-u_0|} \spc {\mbox{for $|u-u_0| \leq \Delta u$}}.
\label{eq:51a}
\end{equation}
Integrating from $u_0$ to $u$ gives
\begin{equation}
\sqrt{\frac{\varepsilon}{u_0}}\: |u-u_0|^{\frac{3}{2}} \;\leq\;
|x(u)| \;\leq\; 2 \:\sqrt{\frac{\varepsilon}{u_0}}\:
|u-u_0|^{\frac{3}{2}} \spc {\mbox{for $|u-u_0| \leq \Delta u$}}.
\label{eq:512}
\end{equation}
Taking the inverses, we obtain for $|u-u_0|$ the bounds
\begin{equation}
\underline{v}(x) \;\leq\; |u(x) - u_0| \;\leq\; \overline{v}(x)
\spc {\mbox{for $|x| \leq \Delta x$,}}    \label{eq:513}
\end{equation}
where we set
\begin{eqnarray}
\underline{v}(x) &=& \left( \frac{x^2\: u_0}{4
\varepsilon}\right)^{\frac{1}{3}} \;,\spc \overline{v}(x) \;=\;
\left( \frac{x^2\: u_0}{\varepsilon}\right)^{\frac{1}{3}}
\label{eq:514} \\
\Delta x &=& \sqrt{\frac{\varepsilon}{u_0}}\: (\Delta
u)^{\frac{3}{2}} \;=\; 2^{-\frac{3}{2}} \:\sqrt{\varepsilon}\: u_0
\;=\; 2^{-\frac{3}{2}} \:\beta \:\varepsilon^{-\frac{1}{2}}\:
(1+{\cal{O}}(\varepsilon))\;. \label{eq:515}
\end{eqnarray}
If $x>\Delta x$, the second inequality in~(\ref{eq:512}) shows that
$u-u_0 \geq 2^{-\frac{5}{3}} u_0$, and thus in this region there
are constants $a_1, a_2>0$ such that
\begin{equation}
a_1\:\sqrt{\varepsilon} \;\leq\; \rho(u) \;\leq\;
a_2\:\sqrt{\varepsilon}\;. \label{eq:51h}
\end{equation}
Hence for $x>\Delta x$,
\[ \frac{1}{a_2\:\sqrt{\varepsilon}} \;\leq\; \frac{du}{dx} \;\leq\;
\frac{1}{a_1\:\sqrt{\varepsilon}} \;. \] Integration shows that,
possibly after decreasing $a_1$ and increasing $a_2$,
\begin{equation}
\underline{u}(x) \;\leq\; u(x) \;\leq\; \overline{u}(x) \spc
{\mbox{for $x>\Delta x$}}    \label{eq:51f}
\end{equation}
with
\begin{equation}
\underline{u}(x) \;=\; \frac{x}{a_2\:\sqrt{\varepsilon}}\;,\spc
\overline{u}(x) \;=\; \frac{x}{a_1\:\sqrt{\varepsilon}}\;.
\label{eq:51g}
\end{equation}
If on the other hand $x<-\Delta x$, we see from~(\ref{eq:512})
that $u-u_0 \leq -2^{-\frac{5}{3}} u_0$, and thus $\rho$ can be
estimated by
\begin{equation}
\frac{b_1}{2}\: u^{-\frac{1}{2}} \;\leq\; \rho(u) \;\leq\;
\frac{b_2}{2}\: u^{-\frac{1}{2}}    \label{eq:51l}
\end{equation}
with $b_1, b_2>0$. We integrate from $\umin$ to $u$,
\[ b_1\:\sqrt{u} - b_1\:\sqrt{\umin} \;\leq\; x(u) - \xmin \;\leq\;
b_2\:\sqrt{u} -
b_2\: \sqrt{\umin}\;, \] and solve for $u$. This gives
\begin{equation}
\underline{u}(x) \;\leq\; u(x) \;\leq\; \overline{u}(x) \spc
{\mbox{for $\xmin \leq x < -\Delta x$}}    \label{eq:51j}
\end{equation}
with
\begin{equation}
\underline{u}(x) \;=\; \left( \frac{x - \xmin}{b_2} + \sqrt{\umin}
\right)^2 \;,\spc \overline{u}(x) \;=\; \left( \frac{x -
\xmin}{b_1} + \sqrt{\umin} \right)^2 \;. \label{eq:51k}
\end{equation}

For any $x_1>0$, we can, by choosing $\delta$ small enough,
arrange that $\Delta x$, (\ref{eq:515}), is greater than $x_1$.
Thus $u(x)$ is on the interval $[-x_1, x_1]$ bounded
by~(\ref{eq:513}). Substituting these bounds into~(\ref{eq:510})
and using~(\ref{eq:509}) gives~(\ref{eq:5nasy}).

To show that $\Gamma(x) h'(x) \geq 0$, note that
\[ h'(x) \;=\; h'(u) \: \frac{du}{dx} \;, \]
where $\frac{du}{dx} = \rho^{-1}$ is positive, and the sign of 
$h'(u)$ is obtained from~(\ref{eq:510}).

For the derivation of the inequalities~(\ref{eq:56})--(\ref{eq:58}),
we consider the three regions $x<-\Delta x$, $|x| \leq \Delta x$, and
$x>\Delta x$ separately.  We begin with the case $|x| \leq \Delta x$. 
For $h'(x)$ and $h''(x)$, we have again the bounds~(\ref{eq:59b})
and~(\ref{eq:51b}), respectively.  Using that $u \geq u_0/2$ as well
as~(\ref{eq:51a}), (\ref{eq:513}), and~(\ref{eq:514}), we obtain
\begin{eqnarray*}
|h'(x)| &\leq& \frac{c}{u^2 \rho^3} \;\leq\; \frac{c}{\sqrt{u_0}\:
\varepsilon^\frac{3}{2}}\: |u-u_0|^{-\frac{3}{2}}
\;\leq\; \frac{c}{u_0 \:\varepsilon\: x} \\
|h''(x)| &\leq& \frac{c}{u^4 \rho^6} + \frac{c}{u^3 \rho^4}
\;\leq\; \frac{c}{u_0^2\:\varepsilon^2\:x^2} +
\frac{c}{u_0^\frac{5}{3}\: \varepsilon^\frac{4}{3}\: x^{\frac{4}{3}}}\;,
\end{eqnarray*}
and in view of~(\ref{eq:509}) and~(\ref{eq:515}), this
implies~(\ref{eq:56}) and~(\ref{eq:57}).  To compute
$h_\varepsilon(x)$, we again use~(\ref{eq:5G}), but now with
\[ s(u) \;=\; \sqrt{\frac{\varepsilon}{u_0}}\: |u-u_0|^{\frac{3}{2}}\:
\Gamma(u-u_0) \spc {\mbox{for $|u-u_0| \leq \Delta u$}} \;, \]
where as before $\Gamma$ is the step function $\Gamma(\tau) =
{\mbox{sgn}}(\tau)$.  The first summand in~(\ref{eq:5G}) is computed
as follows,
\begin{eqnarray}
u(s) &=& u_0 + u_0^{\frac{1}{3}}\: \varepsilon^{-\frac{1}{3}}\:
s^{\frac{2}{3}}\: \Gamma(s) \spc {\mbox{for $|s| \leq \Delta x$}}
\label{eq:40a} \\
&\stackrel{(\ref{eq:509})}{=}& \frac{1}{\varepsilon}\: (\beta +
(\beta \varepsilon s^2)^{\frac{1}{3}}\: \Gamma(s))\:
(1+{\cal{O}}(\varepsilon))
\label{eq:516} \\
h(s) &\stackrel{(\ref{eq:510})}{=}&
\frac{1}{2}\:\log \left( \frac{\beta^{\frac{1}{3}}\:
\varepsilon^{\frac{4}{3}}\: s^{\frac{2}{3}}}{\beta + (\beta
\varepsilon s^2)^{\frac{1}{3}} \:\Gamma(s)}
\left(1+\varepsilon\: {\cal{O}}(s^0) + \varepsilon^{\frac{4}{3}}\:
{\cal{O}}(s^{\frac{2}{3}})
\right) \right) - \log (\omega+m) \nonumber \\
\varepsilon\: h_\varepsilon(s) &=& \left( \frac{2}{3} -
\frac{1}{6}\: \frac{(\beta \varepsilon s^2)^{\frac{1}{3}}
\:\Gamma(s)} {\beta + (\beta \varepsilon s^2)^{\frac{1}{3}}
\:\Gamma(s)} \right) \left(1+\varepsilon \:{\cal{O}}(s^0) +
\varepsilon^{\frac{4}{3}}\: {\cal{O}}(s^{\frac{2}{3}}) \right) \;.
\nonumber
\end{eqnarray}
Differentiating with respect to $s$ gives the bound
\[ \varepsilon\:|h'_\varepsilon(s)| \;\leq\; c\: \varepsilon^{\frac{1}{3}}
\: s^{-\frac{1}{3}} + c\: \varepsilon^{\frac{2}{3}}\:
s^{\frac{1}{3}} \;, \]
and thus
\[ \int_{-\Delta x}^{\Delta x} \varepsilon\:|h'_\varepsilon(s)|\:ds
\;\leq\; c\: \varepsilon^{\frac{1}{3}}\:(\Delta x)^{\frac{2}{3}} +
c\: \varepsilon^{\frac{2}{3}}\: (\Delta x)^{\frac{4}{3}} \;. \]
Using~(\ref{eq:515}), we conclude that the total variation of
$\varepsilon h_\varepsilon(s)$ is bounded uniformly in
$\varepsilon$. In order to estimate the total variation of the
second summand in~(\ref{eq:5G}), we first compute $x'(s)$,
\[ x'(s)^2 \;=\; (\rho(s)\: u'(s))^2 \;=\; \frac{16}{9}\:
\frac{u_0}{u(s)} \left(1+\varepsilon\: {\cal{O}}(s^0) +
\varepsilon^{\frac{4}{3}} \:{\cal{O}}(s^{\frac{2}{3}}) \right) \;.
\]
This is uniformly bounded from above and below, proving~(\ref{eq:5I}).
Differentiating with respect to $\varepsilon$ using~(\ref{eq:516})
gives the estimate
\begin{equation}
\varepsilon\: |x'_\varepsilon(s)| \;\leq\; c\:
\varepsilon^{\frac{1}{3}}\: s^{\frac{2}{3}} \;. \label{eq:517}
\end{equation}
Since $x(s=0)=0$ for all $\varepsilon$ (from~(\ref{eq:40a})),
integration yields that
\begin{eqnarray}
\varepsilon\: |x_\varepsilon(\pm \Delta x)| &\leq& \left|
\int_0^{\pm \Delta x} \varepsilon\:|x'_\varepsilon(s)|\: ds
\right| \;\leq\; c\: \varepsilon^{\frac{1}{3}}\: (\Delta
x)^{\frac{5}{3}} \;\stackrel{(\ref{eq:515})}{\leq}\; c\:
\varepsilon^{-\frac{1}{2}}
\label{eq:51A} \\
\int_{-\Delta x}^{\Delta x} \frac{\varepsilon\:
|x'_\varepsilon(s)|}{s}\: ds &\leq& c\:
\varepsilon^{\frac{1}{3}}\:(\Delta x)^{\frac{2}{3}} \;\leq\;c \;.
\label{eq:51B}
\end{eqnarray}
Furthermore,
\begin{equation}
\int_{-\Delta x}^{\Delta x} \frac{\varepsilon\:
|x_\varepsilon(s)|}{s^2}\: ds \;\leq\; \int_{-\Delta x}^{\Delta x}
\frac{1}{s^2} \left( \int_0^s \varepsilon\: |x'_\varepsilon(t)|\:
dt \right) ds . \label{eq:51c}
\end{equation}
Using~(\ref{eq:517}), the inner integral is for small $s$ bounded
by a constant times $t^{\frac{5}{3}}$. Thus when integrating by
parts, we get no boundary terms at $s=0$ and obtain
\begin{eqnarray}
\int_{-\Delta x}^{\Delta x} \frac{\varepsilon\:
|x_\varepsilon(s)}{s^2}\: ds \;\leq\; \frac{1}{\Delta x}
\int_{-\Delta x}^{\Delta x} \varepsilon\: |x'_\varepsilon(t)|\: dt
+ \int_{-\Delta x}^{\Delta x} \frac{\varepsilon\:
|x'_\varepsilon(t)|}{s}\: ds \;\leq\; c \;, \label{eq:51d}
\end{eqnarray}
where in the last step we used~(\ref{eq:517}) and~(\ref{eq:51B}).
Combining~(\ref{eq:51B}) and~(\ref{eq:51d}) with the
estimates~(\ref{eq:56}) and~(\ref{eq:57}), we conclude that the
total variation of the second summand in~(\ref{eq:5G}) is bounded
uniformly in $\varepsilon$. This shows that~(\ref{eq:58}) holds if
the integration domain is restricted to $x \in (-\Delta x, \Delta
x)$.

In the case $x>\Delta x$, (\ref{eq:51h}), (\ref{eq:51f}), and (\ref{eq:51g})
yield, again using~(\ref{eq:59b}) and~(\ref{eq:51b}),
\begin{eqnarray*}
|h'(x)| &\leq& \frac{c}{u^2 \rho^3} \;\leq\; \frac{c}{x^2\:
\sqrt{\varepsilon}} \;\leq\; \frac{c}{x}\:
\frac{1}{\sqrt{\varepsilon}
\:\Delta x} \;\stackrel{(\ref{eq:515})}{\leq}\; \frac{c}{x} \\
|h''(x)| &\leq& \frac{c}{u^4 \rho^6} + \frac{c}{u^3 \rho^4}
\;\leq\; \frac{c}{x^4\:\varepsilon} + \frac{c}{x^3\:
\sqrt{\varepsilon}} \;\leq\; \frac{c}{x^2}\;,
\end{eqnarray*}
proving~(\ref{eq:56}) and~(\ref{eq:57}). To compute the total
variation of $h_\varepsilon(x)$, we apply~(\ref{eq:5G}) with
\[ s(u) \;=\; \frac{\sqrt{\varepsilon}}{3\:\sqrt{2}}\:u \spc
{\mbox{for $u>u_0+\Delta u$}}. \]
Using that
\[ s(u_0 + \Delta u) \;=\; s\! \left( \frac{3 u_{0}}{2} \right)
\;\stackrel{(\ref{eq:515})}{=} \Delta x\;, \]
we see that
\[ u(s) \;=\; \frac{3\:\sqrt{2}}{\sqrt{\varepsilon}}\:s \spc
{\mbox{for $s>\Delta x$}}. \]
Moreover, from~(\ref{eq:510}) and~(\ref{eq:511}),
\begin{eqnarray*}
h(s) &=& \frac{1}{2}\:\log \left(
\frac{\varepsilon}{(\omega+m)^2}\: \frac{3\:\sqrt{2}\:s - \beta\:
\varepsilon^{-\frac{1}{2}}} {3\:\sqrt{2}\:s} \left( 1 +
\varepsilon\:{\cal{O}}(s^0) +
\sqrt{\varepsilon}\:{\cal{O}}(s^{-1}) \right) \right) \\
\varepsilon\: h_\varepsilon(s) &=& \frac{1}{2} \left(1 + \frac{\beta\:
\varepsilon^{-\frac{1}{2}}}{6\:\sqrt{2}\:s - 2 \beta\:
\varepsilon^{-\frac{1}{2}}} \right) \left( 1 +
\varepsilon\:{\cal{O}}(s^0) +
\sqrt{\varepsilon}\:{\cal{O}}(s^{-1}) \right) \\
\varepsilon\:|h'_\varepsilon(s)| &\leq& \frac{c}{\sqrt{\varepsilon}\:s^2} \\
\int_{\Delta x}^\infty \varepsilon\: |h'_\varepsilon(s)|\:ds
&\leq&
\frac{c}{\sqrt{\varepsilon}\:\Delta x} \;\leq\; c \\
x'(s)^2 &=& (\rho(s)\: u'(s))^2 \;=\; 72 \left( 1 -
\frac{\beta}{3\:\sqrt{2} \: s \: \sqrt{\varepsilon}} \right)
\left(1 + \varepsilon{\cal{O}}(s^0) + \sqrt{\varepsilon}\:
{\cal{O}}(s^{-1}) \right) \;.
\end{eqnarray*}
Since $s>\Delta x$, we conclude from~(\ref{eq:515}) that~(\ref{eq:5I})
holds.  Differentiating the last relation with respect to
$\varepsilon$ and integrating gives
\begin{eqnarray}
\varepsilon\:|x'_\varepsilon(s)| &\leq& \frac{c}{s\:
\sqrt{\varepsilon}}
\nonumber \\
\int_{\Delta x}^\infty \frac{\varepsilon\:|x'_\varepsilon(s)|}{s}
\: ds &\leq& \frac{c}{\sqrt{\varepsilon}\: \Delta x} \;\leq\; c
\label{eq:51i} \\
\int_{\Delta x}^\infty \frac{\varepsilon\:
|x_\varepsilon(s)|}{s^2}\: ds &\leq& \int_{\Delta x}^\infty
\frac{1}{s^2} \left( \int_{\Delta x}^s
\varepsilon\:|x'_\varepsilon(t)|\: dt + \varepsilon\:
x_\varepsilon(\Delta x)
\right) ds \nonumber \\
&\leq& \frac{\varepsilon\: |x_\varepsilon(\Delta x)|}{\Delta x} +
\int_{\Delta x}^\infty \frac{\varepsilon\:
|x_\varepsilon(s)|}{s}\: ds \;\leq\; c\;, \nonumber
\end{eqnarray}
where in the last step we used~(\ref{eq:51A}) and~(\ref{eq:51i}).
This proves~(\ref{eq:58}) if $x>\Delta x$.

\begin{figure}[tb]
    \epsfxsize=16cm
    \centerline{\epsfbox{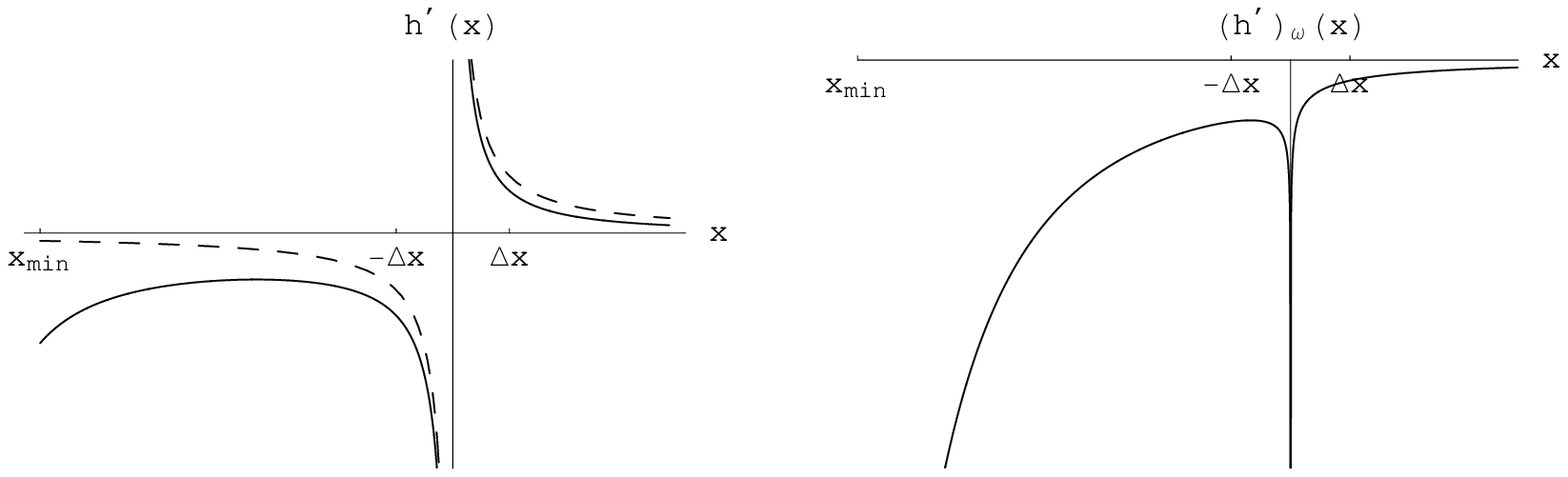}}
    \caption{Typical plots for $h'(x)$ and $h'_\omega(x)$
in the case $\omega<m$.}
    \label{fig3}
\end{figure}

Finally, if $x<-\Delta x$, the bounds~(\ref{eq:51j})
and~(\ref{eq:51k}) give
\begin{eqnarray*}
|h'(x)| &\stackrel{(\ref{eq:59b})}{\leq}&
\frac{c}{u^2 \rho^3} \;\leq\; \frac{8c}{b_1^3}\:
u^{-\frac{1}{2}} \;=\; \frac{c}{b_2\: \sqrt{u}} \\
&\leq& \frac{c}{x-\xmin + b_2\: \sqrt{\umin}}
\;\leq\; \frac{c}{x-\xmin+1} \\
|h''(x)| &\stackrel{(\ref{eq:51b})}{\leq}&
\frac{c}{u^4 \rho^6} + \frac{c}{u^3 \rho^4}
\;\leq\; \frac{c}{u} \;\leq\; \frac{c}{(x-\xmin+1)^2} \;.
\end{eqnarray*}
This concludes the proof of~(\ref{eq:56}) and~(\ref{eq:57}). In order 
to prove~(\ref{eq:58}), we apply~(\ref{eq:5G}) with
\[ s(u) \;=\; \frac{1}{2}\:\sqrt{\beta \:u} - 2 \Delta x \spc
{\mbox{for $\umin \leq u \leq u_0 - \Delta u$}}. \]
Similar as in the case $x>\Delta x$,
\begin{eqnarray}
u(s) &=& \frac{4}{\beta}\: (s+2 \Delta x)^2 \spc {\mbox{for $\smin
\equiv \frac{1}{2} \:\sqrt{\beta \:\umin} - 2 \Delta x
\;\leq\; s \;\leq\; -\Delta x$}} \label{eq:5srange} \\
h(s) &=& \frac{1}{2} \:\log \left( \frac{1}{(\omega+m)^2} \left(
\frac{\beta^2\: (1+{\cal{O}}(\varepsilon)}
{4 (s+2 \Delta x)^2} -\varepsilon \right)\left(1
+ {\cal{O}}((s+2 \Delta x)^{-2}) \right) \right) \nonumber \\
\varepsilon\: h_\varepsilon(s) &=& -\frac{\varepsilon}{2}\:
\frac{4 (s+2 \Delta x)^2 + \beta^2\: 
{\cal{O}}(\varepsilon)}{\beta^2\:(1+{\cal{O}}(\varepsilon)) - 4 
\varepsilon\: (s+2 \Delta x)^2} + \varepsilon\: {\cal{O}}(s^0) +
\frac{\varepsilon\: {\cal{O}}((s+2 \Delta x)^{-2})}
{1+{\cal{O}}((s+2 \Delta x)^{-2})} \nonumber \\
&=&-\frac{\varepsilon}{2}\:
\frac{4 (s+2\Delta x)^2 + \epsilon\:{\cal{O}}(s^0)}
{\beta^2 - 4 \varepsilon\: (s+2 \Delta x)^2} \:+\: \varepsilon\:
{\cal{O}}(s^0) + \varepsilon\: {\cal{O}}((s+2
\Delta x)^{-2}) \nonumber \\
\varepsilon\: |h'_\varepsilon(s)| &\leq& c\:\varepsilon\:(s+2
\Delta x) \;. \nonumber
\end{eqnarray}
Integration yields that
\[ \int_{\ssmin}^{-\Delta x} \varepsilon\: |h'_\varepsilon(s)|\: ds
\;\stackrel{(\ref{eq:5srange})}{\leq} c \varepsilon \left(
\Delta x^2 - \frac{1}{4}\: \beta\: \umin \right)
\;\stackrel{(\ref{eq:515}),(\ref{eq:509})}{\leq};
c \:\varepsilon\: (u_0 - \Delta u - \umin) \;\leq\; c\;. \]
Moreover,
\[ x'(s)^2 \;=\; (\rho(s)\: u'(s))^2 \;=\; 64 \left(1 - \frac{4}{\beta^2}\:
\varepsilon\: (s+2 \Delta x)^2 \right) \left(1 + \varepsilon\:
{\cal{O}}(s^0) + {\cal{O}}((s+2 \Delta x)^{-2}) \right) \;. \]
Using that $(s+2\Delta x)^2 \leq (\Delta x)^2$, we conclude that
$x'(s)$ is uniformly bounded from above and below. We
differentiate with respect to $\varepsilon$ and integrate to
finally obtain similar to~(\ref{eq:5I})--(\ref{eq:5J}),
\begin{eqnarray}
\varepsilon\: |x'_\varepsilon| &\leq& c \varepsilon\: (s+2 \Delta
x)^2
\label{eq:52a} \\
\int_{\ssmin}^{-\Delta x} \frac{\varepsilon\: |x'_\varepsilon(s)|}
{s+2 \Delta x} \:ds &\leq& c \varepsilon\: (u_0 - \Delta u -
\umin)
\;\leq\; c \label{eq:52b} \\
\int_{\ssmin}^{-\Delta x} \frac{\varepsilon\: |x_\varepsilon(s)|}
{(s+2 \Delta x)^2} \:ds &\leq& \int_{\ssmin}^{-\Delta x}
\frac{1}{(s+2 \Delta x)^2} \left( \int_{-\Delta x}^s \varepsilon\:
|x'_\varepsilon(t)|\:dt + \varepsilon\:
|x_\varepsilon(-\Delta x)| \right) ds \nonumber \\
&&\hspace*{-3cm} \leq\; \frac{\varepsilon\: |x_\varepsilon(-\Delta
x)|}{\Delta x} + \frac{1}{\smin + 2 \Delta x}
\int_{\ssmin}^{-\Delta x} \varepsilon\: |x'_\varepsilon(t)|\:dt +
\int_{\ssmin}^{-\Delta x} \frac{\varepsilon\:
|x'_\varepsilon(s)|}{s+2 \Delta x}\: ds \;\leq\; c\;, \nonumber
\end{eqnarray}
where in the last line we integrated by parts and used~(\ref{eq:51A})
and~(\ref{eq:52a}),(\ref{eq:52b}). This yields~(\ref{eq:58}) and 
completes the proof of Lemma~\ref{lemma52}.
\QED
The above estimates are illustrated in Figure~\ref{fig3}.
The dashed curve is the graph of $(3x)^{-1}$; it is the
$x$-derivative of the asymptotic function $\frac{1}{3} \log
(\varepsilon^2 x)$ which appears in~(\ref{eq:5nasy}).

The next lemma controls the behavior of $\xmin$.
\begin{Lemma} \label{lemma52a}
There are constants $c, \delta>0$ such that for all $\omega \in
(m-\delta, m)$,
\begin{eqnarray}
\left| \xmin + 4 \beta\: (m^2 - \omega^2)^{-\frac{1}{2}} \right|
&\leq& c
\label{eq:5xx} \\
\left| \partial_\omega \left(\xmin + 4 \beta\: (m^2 -
\omega^2)^{-\frac{1}{2}} \right) \right| &\leq&
\frac{c}{m-\omega}\;. \label{eq:5yy}
\end{eqnarray}
\end{Lemma}
{\Proof} According to the definition of $x$ and~(\ref{eq:511}),
\begin{equation}
\xmin \;=\; -2 \int_{\sumin}^{u_0} \rho(u)\: du
\;=\; -2 \int_{\sumin}^{u_0} \sqrt{\varepsilon\:
\frac{|u-u_0|}{u} \left(1+{\cal{O}}(u^{-1}) \right)}\: du \;,
\label{eq:5zz}
\end{equation}
and a calculation using~(\ref{eq:509}) shows that the leading
contribution in $\sqrt{\varepsilon}$ to this integral is $4 \beta
\varepsilon^{-\frac{1}{2}}$ (this can be readily verified using
Mathematica). This proves~(\ref{eq:5xx}).
Differentiating~(\ref{eq:5zz}) with respect to $\omega$ and
estimating the resulting integral gives~(\ref{eq:5yy}). \QED

\section{The Region $\omega>m$} \label{sec6}
\setcounter{equation}{0}
We turn now to the planar equation~(\ref{eq:2y}).  Consider the two
solutions $(\vartheta^{(b)}, L^{(b)})$, $b=1,2$, with
\begin{equation}
\lim_{x \rightarrow -\infty} (\vartheta^{(1)}, L^{(1)}) \;=\;
(0,0)\;\;\;,\spc \lim_{x \rightarrow -\infty} (\vartheta^{(2)},
L^{(2)}) \;=\; (\pi,0) \label{eq:3d}
\end{equation}
and define in analogy to the transmission coefficients, the
quantities
$(\vartheta^{(b)}_\infty, L^{(b)}_\infty)$ by
\begin{equation}
(\vartheta^{(b)}_\infty, L^{(b)}_\infty) \;=\; \lim_{x \rightarrow
\infty} (\vartheta^{(b)}(x), L^{(b)}(x))\;. \label{eq:3g}
\end{equation}
The next lemma expresses the coefficients $t_{ab}$ in the integral
representation~(\ref{eq:1a}) in terms of these ``transmission
coefficients.''
\begin{Lemma} \label{lemma31}
The coefficients $t_{ab}$, (\ref{eq:k}), are for $\omega>m$ given
by
\begin{eqnarray}
t_{11} &=& t_{22} \;=\; \frac{1}{2} \label{eq:13a} \\
t_{12} &=& \overline{t_{21}} \;=\; \frac{1}{2}\: e^{-i \beta_0}\:
\tanh z \;, \label{eq:13b}
\end{eqnarray}
where
\begin{eqnarray}
\beta_0 &=& \arctan \frac{\lambda}{m r_1} \\
z &=& \frac{L^{(1)}_\infty - L^{(2)}_\infty}{4} \:+\: i\:
\frac{\vartheta^{(1)}_\infty - \vartheta^{(2)}_\infty + \pi}{4}
\;. \label{eq:6zf}
\end{eqnarray}
\end{Lemma}
{\Proof} According to~(\ref{eq:25a}), (\ref{eq:2phi}),
and~(\ref{eq:19a}), the functions $(\vartheta, L)$
correspond to the $2$-spinor
\begin{equation}
X \;=\; e^{-\frac{L}{2}} \left( \begin{array}{ll}
e^{-\frac{i \beta}{2}} & 0 \\ 0 & e^{\frac{i \beta}{2}}
\end{array} \right)
\left( \begin{array}{cc} e^{\frac{h}{2}}\: \cos
\frac{x+\vartheta}{2}
\:-\: i\:  e^{-\frac{h}{2}}\: \sin \frac{x+\vartheta}{2} \\
-e^{\frac{h}{2}}\: \cos \frac{x+\vartheta}{2} \:-\: i\:
e^{-\frac{h}{2}}\: \sin \frac{x+\vartheta}{2} \end{array} \right)
\;. \label{eq:3e}
\end{equation}
In the limit $u \rightarrow -\infty$, the function $h \rightarrow
0$ (cf.\ (\ref{eq:24a}) and observe that $\lim_{r \searrow r_1} f(r) =
\omega
= \lim_{r \searrow r_1} g(r)$ according to~(\ref{eq:23a})
and~(\ref{eq:2xx}),(\ref{eq:2w})), and $x$ coincides asymptotically with
$\Omega_0 u$, up to an  (irrelevant) additive constant. Thus
comparing~(\ref{eq:3e}) with~(\ref{eq:c}) and using~(\ref{eq:3d}) gives
\[ f_0^{(1)} \;=\; \left( \begin{array}{c} e^{-\frac{i \beta_0}{2}} \\
-e^{\frac{i \beta_0}{2}} \end{array} \right) \;\;\;,\spc
f_0^{(2)} \;=\; \left( \begin{array}{c} -i e^{-\frac{i \beta_0}{2}} \\
-i e^{\frac{i \beta_0}{2}} \end{array} \right) \;. \] Hence the
fundamental solutions $X_0$ and $X_1$, which are characterized
by~(\ref{eq:d}), are the linear combinations
\[ X_1 \;=\; \frac{1}{2}\: e^{\frac{i \beta_0}{2}} \:(X^{(1)} + i X^{(2)})
\;\;\;,\spc X_2 \;=\; \frac{1}{2}\: e^{-\frac{i \beta_0}{2}}
\:(-X^{(1)} + i X^{(2)}) \;. \] We next consider~(\ref{eq:3e})
in the limit $u \rightarrow +\infty$. According to~(\ref{eq:24a}),
(\ref{eq:23a}), (\ref{eq:2xx})--(\ref{eq:2w}), and~(\ref{eq:14a}),
\[ \lim_{x \rightarrow \infty} h(x) \;=\; \frac{1}{2}\:\log
\frac{\omega - m}{\omega + m} \;=\; -2 \Theta \;. \]
Also, $\beta$ goes to zero in this limit. Hence using~(\ref{eq:b})
and~(\ref{eq:3g}), one sees that
\[ f_{\infty\:1\!/\!2} \;=\; \frac{1}{2}\: e^{\pm \frac{i \beta_0}{2}}
\left( \displaystyle \begin{array}{c} \displaystyle \pm \exp
\left(-\frac{L^{(1)}_\infty + i \vartheta^{(1)}_\infty}{2} \right)
+ i\: \exp \left(-\frac{L^{(2)}_\infty + i
\vartheta^{(2)}_\infty}{2} \right)
\\[1.1em]
\displaystyle \pm \exp \left(-\frac{L^{(1)}_\infty - i
\vartheta^{(1)}_\infty}{2} \right) + i\: \exp
\left(-\frac{L^{(2)}_\infty - i \vartheta^{(2)}_\infty}{2} \right)
\end{array} \right) \;. \]
Substituting this last formula into~(\ref{eq:j}) yields
\[ t_{1\!/\!2}(\alpha) \;=\; e^{\mp \frac{i \beta_0}{2}} \left[ i
e^{-\frac{L^{(1)}_\infty}{2}}\: \sin \left( \alpha +
\frac{\vartheta^{(1)}_\infty}{2} \right) \pm
e^{-\frac{L^{(2)}_\infty}{2}}\: \sin \left( \alpha +
\frac{\vartheta^{(2)}_\infty}{2} \right) \right] \;. \]
A short
calculation shows that
\begin{equation}
|t_1|^2 \;=\; |t_2|^2 \;. \label{eq:n}
\end{equation}
Together with~(\ref{eq:k}), this immediately
yields~(\ref{eq:13a}). Furthermore, it is obvious
from~(\ref{eq:k}) that $\overline{t_{12}}=t_{21}$. Thus it remains
to compute $t_{12}$. According to~(\ref{eq:k}) and~(\ref{eq:n}),
we have
\begin{eqnarray*}
t_{12} &=& \frac{1}{2 \pi} \:\int_0^{2 \pi} \frac{t_1
\:\overline{t_2}}{2 \:|t_2|^2} \;=\; \frac{1}{4 \pi} \:\int_0^{2
\pi}
\frac{t_1}{t_2} \\
&=& \frac{e^{-i \beta_0}}{4 \pi} \int_0^{2 \pi} \frac{i \rho_1\:
\sin (\alpha+\varphi_1) + \rho_2 \:\sin (\alpha+\varphi_2)} {i
\rho_1\: \sin (\alpha+\varphi_1) - \rho_2 \:\sin
(\alpha+\varphi_2)}\: d\alpha \;,
\end{eqnarray*}
where we set $\rho_i = L^{(i)}_\infty/2$ and $\varphi_i =
\vartheta^{(i)}_\infty/2$. It is convenient to shift the integration
variable by $\alpha \rightarrow \alpha - \varphi_2$ and to divide
the numerator and denominator by $\rho_2$. This gives
\[ t_{12} \;=\; \frac{e^{-i \beta}}{4 \pi} \int_0^{2 \pi}
\frac{i \rho\: \sin (\alpha+\varphi) + \sin \alpha} {i \rho\: \sin
(\alpha+\varphi) - \sin \alpha}\: d\alpha \] with $\rho
\equiv\rho_1/\rho_2$ and $\varphi \equiv \varphi_1 - \varphi_2$.
We express the trigonometric functions as exponentials and set
$\mu = \rho e^{-i \varphi}$,
\[ t_{12} \;=\; \frac{e^{-i \beta_0}}{4 \pi} \int_0^{2 \pi}
\frac{(\overline{\mu} - i)\: e^{2 i \alpha} - (\mu - i)}
{(\overline{\mu} + i)\: e^{2 i \alpha} - (\mu + i)} \: d\alpha \;.
\]
Setting $z=e^{2 i \alpha}$, the $\alpha$-integral can be regarded
as an integral along the unit circle in the complex plane; more
precisely,
\begin{equation}
t_{12} \;=\; \frac{e^{-i \beta_0}}{4 \pi i} \oint_{|z|=1}
\frac{(\overline{\mu} - i)\: z - (\mu - i)} {(\overline{\mu} +
i)\: z - (\mu + i)}\; \frac{dz}{z}\;.\label{eq:3a}
\end{equation}
This contour integral can be computed with residues as follows.
According to~(\ref{eq:3d}), $\lim_{x \rightarrow -\infty}
\vartheta^{(1)}_\infty - \vartheta^{(1)}_\infty = -\pi$. A
comparison argument using the differential equation for
$\vartheta$, (\ref{eq:2y}), shows that $\vartheta^{(1)}
- \vartheta^{(2)}$ takes values in the interval $(-2 \pi, 0)$
for all $x$. Hence $-\pi < \varphi < 0$, or equivalently,
\[ {\mbox{Im}}\: \mu > \;0 \;. \]
As a consequence, the integrand in~(\ref{eq:3a}) has only one pole
in the unit circle, at $z=0$. We conclude that
\[ t_{12} \;=\; \frac{1}{2}\: e^{-i \beta_0}\: \frac{\mu - i }{\mu + i}
\;=\; \frac{1}{2}\: e^{-i \beta_0} \: \frac{\sqrt{\mu/i} -
\sqrt{i/\mu}} {\sqrt{\mu/i} + \sqrt{i/\mu}} \;, \] and this
coincides with~(\ref{eq:13b}). \QED
The following two lemmas control the behavior of $(\vartheta, L)$
for large $x$.
\begin{Lemma} \label{lemma61}
There is $c>0$ such that for all $\omega \in (m,m+\delta)$ and $x
\in (0, \infty]$,
\begin{equation}
\left| \vartheta(x) - \vartheta(0) \right| \;\leq\; c \;,\spc
\left| L(x) - L(0) \right| \;\leq\; c\;. \label{eq:3b} \\
\end{equation}
\end{Lemma}
{\Proof} According to~(\ref{eq:5a}), there is $x_0>0$ such that
\begin{equation}
1 + h'\: \sin (x+\vartheta) \;>\; \frac{1}{2} \spc {\mbox{for all
$x>x_0$.}}\label{eq:3dd}
\end{equation}
On the interval $[0,x_0]$, we can control $\vartheta$ by
integrating the
$\vartheta$-equation in~(\ref{eq:2y}),
\begin{equation}
|\vartheta(x_0) - \vartheta(0)| \;=\; \left| \int_0^{x_0} h'\:
\sin(x+\vartheta)\: dx \right| \;\stackrel{(\ref{eq:5a})}{=}\;
c\:x_0\;.\label{eq:6NZ}
\end{equation}
In the region $x>x_0$, we again integrate the equation,
\[ \vartheta(x) - \vartheta(x_0) \;=\; \int_{x_0}^x h'\: \sin (x+\vartheta)
\:dx \;=\;
-\int_{x_0}^x \frac{h'}{1+h' \sin (\tau+\vartheta)}\:
\frac{d}{d\tau}(\cos (\tau+\vartheta))\: d\tau\;. \]
We integrate by parts and, using~(\ref{eq:3dd}) and 
Lemma~\ref{lemma51}, we find
\begin{equation}
|\vartheta(x) - \vartheta(x_0)| \;\leq\; 2\: (|h'(x)| + |h'(x_0)|)
+ 4 \int_{x_0}^x \left( |h''| + h'^2 \right) d\tau \;\leq\; c\;.
\label{eq:6NY}
\end{equation}
The second statement in~(\ref{eq:3b}) follows similarly by integrating
the $L$-equation in~(\ref{eq:2y}).
\QED

\begin{Lemma} \label{lemma62}
There is $c>0$ such that for all $\omega \in (m,m+\delta)$ and $x
\in (0, \infty]$,
\begin{equation}
|\vartheta_\omega(x) - \vartheta_\omega(0)| \;\leq\;
\frac{c}{\omega-m}\;,\spc |L_\omega(x) - L_\omega(0)| \;\leq\;
\frac{c}{\omega-m} \;. \label{eq:6a}
\end{equation}
\end{Lemma}
{\Proof} Differentiating through the ODEs in~(\ref{eq:2y}) with
respect to $\omega$ gives
\begin{eqnarray}
\vartheta_\omega' &=& h'\: \cos(x+\vartheta) \:\vartheta_\omega
\:+\:
h'_\omega\: \sin(x+\vartheta) \nonumber \\
&=& L'\:\vartheta_\omega +
h'_\omega\:\sin(x+\vartheta) \label{eq:3k} \\
L_\omega' &=& -h'\: \sin(x+\vartheta) \:\vartheta_\omega \:+\:
h'_\omega\: \cos(x+\vartheta) \label{eq:3l} \;.
\end{eqnarray}
The differential equation~(\ref{eq:3k}) can be solved using the
method of variation of constants. The solution is
\begin{equation}
\vartheta_\omega(x) - \vartheta_\omega(0) \;=\; e^{L(x)} \int_0^x
e^{-L(\tau)}\: h'_\omega(\tau)\: \sin(\tau+\vartheta)\: d\tau \;.
\label{eq:12a}
\end{equation}
Lemma~\ref{lemma61} yields that
\[ |\vartheta_\omega(x) - \vartheta_\omega(0)| \;\leq\; \int_0^x
|h'_\omega(\tau)|\: d\tau\;, \] and the estimate~(\ref{eq:5c}) in
Lemma~\ref{lemma51} gives the first part of~(\ref{eq:6a}). To
derive the second part, we integrate~(\ref{eq:3l}) and apply again
the integration-by-parts technique of Lemma~\ref{lemma61},

\begin{eqnarray}
\lefteqn{ |L_\omega(x) - L_\omega(0)| \;\leq\; \left| \int_0^x
\frac{h'\: \vartheta_\omega}{1+h'\: \sin (\tau+\vartheta)}\left(
\frac{d}{d \tau}\: \cos(\tau+\vartheta) \right) d\tau \:+\:
\int_0^x h'_\omega\:
\cos(\tau+\vartheta)\:d\tau \right| } \nonumber \\
&\leq& 2\: |h'\: \vartheta_\omega|(x) + 2\: |h'\:
\vartheta_\omega|(0) \:+\: 2 \int_0^x \left(|h''\:
\vartheta_\omega| + h'^2 \:|\vartheta_\omega| + |h'\:
\vartheta'_\omega| \right) \:+\: \int_0^x |h'_\omega| \:.\spc\;\;
\label{eq:6b}
\end{eqnarray}
Using the estimates of Lemma~\ref{lemma51} and
Lemma~\ref{lemma61}, the only problematic term is the integral
of $|h'\: \vartheta'_\omega|$. But from~(\ref{eq:3k})
and~(\ref{eq:2y}) we have
\begin{eqnarray*}
|h'\: \vartheta'_\omega| &=& |h'^2 \: \sin(x+\vartheta)\:
\vartheta_\omega +
h'\: h'_\omega\: \sin(x+\vartheta)| \\
&\leq& |h'^2\: \vartheta_\omega| + |h'\: h'_\omega| \;\leq\;
\frac{1}{\omega-m}\: \frac{c}{(1+x)^2} \;,
\end{eqnarray*}
where in the last step we used~(\ref{eq:5a}) and the first part
of~(\ref{eq:6a}). \QED
We remark that by combining~(\ref{eq:3l})
and~(\ref{eq:2y}), we can write the $L_\omega$-equation as
\[ L'_\omega \;=\; -\vartheta'\: \vartheta_\omega + h'_\omega\:
\cos(x+\vartheta)\;. \] Although this looks very similar
to~(\ref{eq:3k}), it seems difficult to deduce the second part
of~(\ref{eq:6a}) by integration (note that the total variation
of $\vartheta$ need not be bounded uniformly in $\omega$). This is
the reason why we instead used an integration-by-parts argument.

We are now in the position to prove that in the integral
representation~(\ref{eq:1b}), the contributions for $\omega>m$
decay in $t$ at least at the rate $t^{-\frac{5}{6}-\varepsilon}$. Consider
the two fundamental solutions $(\vartheta^{(b)}, L^{(b)})$, (\ref{eq:3d}).
For
negative $x$, the function $h'(x)$ is smooth in $\omega$.
Furthermore, $h'(x)$ is computed to be
\begin{equation}
h'(x) \;=\; h'(r) \: \frac{dr}{du} \:\frac{du}{dx} \;
\stackrel{(\ref{eq:2ut})}{=}\; 
h'(r) \: \frac{\Delta}{r^2+a^2} \:\frac{du}{dx} \;.
\label{eq:6hp}
\end{equation}
Using that $\Delta$ decays exponentially as $u \rightarrow -\infty$, and
that
for large negative $x$, $u'(x)$ is bounded away from zero, we see that
$h'(x)$ decays rapidly as $x \rightarrow -\infty$, locally
uniformly in $\omega$. Thus standard Gronwall estimates applied to
the differential equations~(\ref{eq:2y}) yield that
$(\vartheta^{(b)}(0), L^{(b)}(0))$ depends smoothly on $\omega$.
Hence Lemma~\ref{lemma61} and Lemma~\ref{lemma62} give us
information on the transmission coefficients, namely
\begin{equation}
|\vartheta^{(b)}_\infty|, |L^{(b)}_\infty| \;\leq\; c
\spc{\mbox{and}}\spc |\partial_\omega \vartheta^{(b)}_\infty|,
|\partial_\omega L^{(b)}_\infty| \;\leq\; \frac{c}{\omega-m}\;.
\label{eq:6c}
\end{equation}
Next we consider the propagator~(\ref{eq:1b}) for $x$ in a compact
set $K$ and $\Psi_0$ with compact support. Again, standard
Gronwall estimates starting from the event horizon yield that the
fundamental solutions $\Psi^{k \omega n}_a(x)$ depend smoothly on
$\omega$, uniformly for $x \in K$. Hence the only non-smooth terms
are the coefficients $t^{k \omega n}_{ab}$. According to
Lemma~\ref{lemma41}, these coefficients have the same regularity
as the transmission coefficients, (\ref{eq:6c}). Furthermore,
Theorem~\ref{thm21} allows us to restrict attention to a
neighborhood of $\omega=m$, and thus we may assume that the square
bracket in~(\ref{eq:1b}) has compact support. We conclude that
this square bracket satisfies the assumption of
Lemma~\ref{lemma40} (with $\alpha=\omega-m$), and thus its Fourier
transform decays like $t^{-\frac{5}{6}-\varepsilon}$.

\section{The Region $\omega<m$} \label{sec7}
\setcounter{equation}{0}
For $\omega<m$, the coefficients $t_{ab}$ in the integral
representation~(\ref{eq:1b}) have the simple explicit
form~(\ref{eq:k}), and thus our task is to analyze the
$\omega$-dependence of $\Psi^{k \omega n}_1(x)$. We again work in
the variables $(\vartheta, L)$ and set
\[ \phi(x) \;=\; x + \vartheta(x) \;. \]
Recall that $\Psi^{k \omega n}_1$ is the fundamental solution with
exponential
decay at infinity. The following lemma shows that this implies that $\lim_{x
\rightarrow \infty} \phi(x) = -\infty$.

\begin{Lemma} \label{lemma71}
There is a constant $C$ independent of $\omega$ such that for all $x>0$,
\begin{equation}
\phi(x) \;<\; C - \log x \;.    \label{eq:7A}
\end{equation}
\end{Lemma}
{\Proof} Using the bounds~(\ref{eq:56}) and that $h'$ is positive,
we have
\begin{equation}
\phi' \;\geq\; 1 - \frac{c}{2x}\: e^{-\phi} \label{eq:7B}
\end{equation}
with $c$ independent of $\omega$. Suppose that~(\ref{eq:7A}) were
false for some $x=x_0$ and $C=\log c$. Then~(\ref{eq:7B}) implies
that
\begin{equation}
\phi'(x) \;\geq\; \frac{1}{2}\;.    \label{eq:7C}
\end{equation}
Hence at $x$, $\phi$ is monotone increasing, whereas the right
side of~(\ref{eq:7A}) is monotone decreasing. As a
consequence~(\ref{eq:7A}) is violated on an open interval $(x,
x+\epsilon)$. Furthermore, by continuity (\ref{eq:7A}) is violated
on a closed set. We conclude that~(\ref{eq:7A}) is violated for
all $x \in [x_0, \infty)$. This means that~(\ref{eq:7C}) holds for
all $x \geq x_0$, and integration yields that
\begin{equation}
\lim_{x \rightarrow \infty} \phi(x) \;=\; \infty\;.
\label{eq:7D}
\end{equation}
To finish the proof, we shall show that~(\ref{eq:7D}) implies that
the corresponding two-spinor $\psi$, (\ref{eq:an2}), grows
exponentially at infinity, giving the desired contradiction
(note that since $\Psi^{k \omega n}_1$ decays at infinity, $X$, 
$\tilde{X}$, and $\psi$ also vanish at infinity, 
see~(\ref{eq:323}), (\ref{eq:19a}) and~(\ref{eq:7phi})).
According to~(\ref{eq:7D}), $\psi$ behaves for large $x$
asymptotically as
\[ \psi \;=\; e^{\frac{\phi - L}{2}} \left( \begin{array}{c}
e^{\frac{h}{2}} \\
e^{-\frac{h}{2}} \end{array} \right) (1+{\cal{O}}(e^{-2 \phi}))
\;. \]
Furthermore, using~(\ref{eq:2y}),
\[ (\phi - L)' \;=\; 1 + h'\: (\sinh \phi - \cosh \phi) \;=\; 1 - h' \:
e^{-\phi} \;\stackrel{(\ref{eq:5a})}{=}\;
1 + {\cal{O}}(e^{-\phi}) \;. \]
Hence for large $x$, $\phi - L \sim x$, and so $\psi$ grows
asymptotically like $\psi \sim e^x$.
\QED
The inequality~(\ref{eq:7A}) shows in particular that
\begin{equation}
\phi(x) \;<\; -\frac{1}{2} \spc {\mbox{for all $x \geq x_1$}}
\label{eq:7E}
\end{equation}
with $x_1 = \exp(C+\frac{1}{2})$. We next show that $\phi$ leaves
the region
$\{ \phi<-\frac{1}{2} \}$ for positive $x$.
\begin{Lemma} \label{lemma72}
There is $x_0 \geq \nu>0$ with $\nu$ independent of $\omega$ such
that
\begin{equation}
\phi(x_0) \;\geq\; -\frac{1}{2} \;.    \label{eq:7F}
\end{equation}
\end{Lemma}
{\Proof} We introduce for $x>0$ the function
\[ \underline{\vartheta}(x) \;=\; \log \frac{1}{4} \int_x^\infty h'(\tau)\:
e^{-\tau}\: d\tau\;. \]
Since by~(\ref{eq:56}) the integrand is positive,
$\underline{\vartheta}$ is monotone decreasing. According
to~(\ref{eq:56}) and~(\ref{eq:5nasy}),
\[ \lim_{x \rightarrow \infty} \underline{\vartheta}(x) = -\infty\;,\spc
\lim_{x \rightarrow 0} \underline{\vartheta}(x) = \infty\;, \] and
so there is a unique $x_0$ with
\[ \underline{\vartheta}(x_0) \;=\; -\frac{1}{2}\;. \]
Now, choosing $0<y<z$, we have
\[ \int_y^z h'(\tau)\: e^{-\tau}\: d\tau \;\geq\; e^{-z}\:(h(z) - 
h(y))\;. \]
Using~(\ref{eq:5nasy}), we see that for small $y$,
\[ \int_y^\infty h'(\tau)\: e^{-\tau}\: d\tau \;>\; 4\;, \]
implying that $x_0$ is bounded away from zero, uniformly in $\omega$.

We shall now prove that $\underline{\vartheta}(x)$ is a lower bound
for $\vartheta$, i.e.
\begin{equation}
\vartheta(x) \;>\; \underline{\vartheta}(x) \spc {\mbox{for all $x
\geq x_0$}}. \label{eq:7G}
\end{equation}
Thus in the region $x \geq x_1$, we apply~(\ref{eq:7E}) to get the
estimate
\[ \vartheta' \;=\; h'\: \sinh \phi \;<\; -\frac{1}{4}\: h'\:
e^{-\phi} \;=\; -\frac{1}{4}\: h'\: e^{-(x+\vartheta)} \;.  \]
We separate variables,
\[ (e^\vartheta)'(x) \;<\; -\frac{1}{4}\: h'(x)\: e^{-x} ,\, \]
and integrating (using that $e^{\vartheta(\infty)}=0$), we find
\[ e^{\vartheta(x)} \;>\; \frac{1}{4}\: \int_x^\infty h'(\tau)\: e^{-\tau}\:
d\tau\;. \]
Thus $\underline{\vartheta}$ is indeed a lower bound in the region
$x \geq x_1$.

It remains to show that $\vartheta>\underline{\vartheta}$ on the
interval $[x_0, x_1]$. Let us assume the contrary. Then $\vartheta$ 
and $\underline{\vartheta}$ meet somewhere on this interval. Let
\[ y \;=\; \sup_{[x_0, x_1]} \{ x \:|\: \vartheta(x) =
\underline{\vartheta}(x) \}\;.  \]
Then $\vartheta(y) = \underline{\vartheta}(y) < -\frac{1}{2}$, and
thus
\[ \vartheta'(y) \;<\; -\frac{1}{4}\: h'\: e^{-(x+\vartheta)} \;=\;
-\frac{1}{4}\: h'\; e^{-(x+\underline{\vartheta})} \;=\;
\underline{\vartheta}'(y)\;. \]
This contradicts the fact that $\vartheta(x) >
\underline{\vartheta}(x)$ for all $x>y$.
\QED

The next lemma controls the behavior of $\phi$ near the origin and
``matches'' the solution across the singularity at $x=0$.
\begin{Lemma} \label{lemma73}
Suppose that for given $\kappa_1 \leq 0$ and $\kappa_2>0$,
\begin{equation}
-\frac{1}{2} \;\leq\; \phi(x) \;\leq\; 0 \spc {\mbox{for all $x
\in (\kappa_1, \kappa_2)$}}. \label{eq:7b}
\end{equation}
Then there is $\tilde{\kappa}_1$ with $\tilde{\kappa}_1<0$,
$\tilde{\kappa}_1 \leq \kappa_1$ and a
parameter $\lambda \geq 0$ such that
\begin{equation}
\left\{ \begin{array}{rccclcl} \displaystyle
-\lambda\:e^{h(x)}+r(x)
&<& \phi(x) &<& -\lambda\: e^{h(x)} &
\;\;\;\;\;& {\mbox{for $\tilde{\kappa}_1<x<0$}} \\[.5em]
\displaystyle -\lambda \:e^{h(x)} &<& \phi(x) &<& -\lambda\:
e^{h(x)}+r(x) & &{\mbox{for $0<x<\kappa_2$}} \end{array} \right.
\label{eq:72r}
\end{equation}
with
\begin{equation}
r(x) \;=\; e^{h(x)} \:\int_0^x e^{-h(\tau)}\:d\tau \;.
\label{eq:7h}
\end{equation}
\end{Lemma}
Note that the function $r(x)$, (\ref{eq:7h}) is finite according
to~(\ref{eq:5nasy}). \\[.3em]
{\em{Proof of Lemma~\ref{lemma73}. }}
Let us first assume that $\kappa_1<0$.
We set $\tilde{\kappa}_1 = \kappa_1$. We choose
$\eta$ with $0<\eta<\min(-\kappa_1, \kappa_2)$. For
negative $x$, $\phi$ satisfies according to~(\ref{eq:2y}), the
equation $\phi' = 1+h' \sin \phi$. Using the bounds~(\ref{eq:7b})
as well as the fact that $h'$ is negative, we obtain that
\begin{equation}
h'\: (\phi + \phi^2) \;<\; \phi' \;<\; 1+h'\: \phi \label{eq:7c}
\end{equation}
for all $x \in (\kappa_1, 0)$. We choose $x$ in the interval
$(\kappa_1, -\eta)$ and consider the inequality~(\ref{eq:7c})
on the interval $(x, -\eta)$. The inequality on the lhs
can be solved by separation of variables and the rhs by variation
of constants. This gives the explicit bounds
\begin{equation}
e^{h(x) - h(-\eta)}\: \phi(-\eta) + e^{h(x)}\:
\int_x^{-\eta} e^{-h(\tau)} \:d\tau \;<\; \phi(x) \;<\;
\frac{\alpha}{1-\alpha} \spc {\mbox{for $-\kappa<x<-\eta$}}
\label{eq:7d}
\end{equation}
with
\begin{equation}
\alpha \;=\; e^{h(x) - h(-\eta)}\:
\frac{\phi(-\eta)}{1+\phi(-\eta)}\;. \label{eq:7alpha}
\end{equation}
If $x$ is positive, then according to~(\ref{eq:2x}),
$\phi$ satisfies the equation $\phi' = 1+h'\: \sinh \phi$.
Using~(\ref{eq:7b}) and that $h'$ is now positive, we get the bounds
\begin{equation}
-1-h'\: \phi \;<\; -\phi' \;<\; -h'\: (\phi-\phi^2) \spc {\mbox{on
$(0,x)$}}. \label{eq:75x}
\end{equation}
We choose $x$ in the interval $(\eta, \kappa)$ and integrate
these bounds from $\eta$ to $x$. This gives the bounds
\begin{equation}
\frac{\beta}{1+\beta} \;<\; \phi(x) \;<\; e^{h(x) - h(\eta)}
\:\phi(\eta) + e^{h(x)}\:\int_\eta^x\: e^{-h(\tau)}\:
d\tau \spc {\mbox{for $\eta<x<\kappa$}} \label{eq:7e}
\end{equation}
with
\[ \beta \;=\; e^{h(x) - h(\eta)}\:
\frac{\phi(\eta)}{1-\phi(\eta)}\;. \]

We now show that
\begin{equation}
\lim_{\eta \searrow 0} \phi(-\eta) \;=\; 0 \;=\;
\lim_{\eta \searrow 0} \phi(\eta) \;. \label{eq:7epsilon}
\end{equation}
Consider $\phi(-\eta)$. From~(\ref{eq:7d}) and~(\ref{eq:7b}) we 
have for fixed $x$ in the interval $-\kappa < x < -\eta$,
\begin{equation}
-\frac{1}{2} \;\leq\; \phi(x) \;<\; \frac{\alpha}{1-\alpha} \;<\; 0 
\;, \label{eq:7phi}
\end{equation}
and thus there is some $\alpha_0<0$ for which $\alpha>\alpha_0$ if $\eta$
is sufficiently small. According to~(\ref{eq:5nasy}), the 
factor $e^{h(x) - h(-\eta)}$ in~(\ref{eq:7alpha}) tends to $+\infty$ as
$\eta \searrow 0$. We conclude from~(\ref{eq:7alpha}) that
\[ \lim_{\eta \searrow 0} 
\frac{\phi(-\eta)}{1+\phi(-\eta)} \;=\; 0 \;, \]
implying the lhs of~(\ref{eq:7epsilon}). A similar argument using the 
rhs of~(\ref{eq:7e}) gives the rhs of~(\ref{eq:7epsilon}).

Since the planar equation~(\ref{eq:2a}) has smooth coefficients,
it is obvious that $\psi(u)$ is smooth and non-zero. Using the 
ansatz'~(\ref{eq:25a}) and~(\ref{eq:an2}) as well 
as~(\ref{eq:7epsilon}), we see that the following limits exist,
\[ \lim_{\eta \searrow 0} e^{-\frac{L(-\eta) - h(-\eta)}{2}}
\left( \begin{array}{c} 1 \\ \displaystyle e^{-h(-\eta)} \: 
\frac{\phi(-\eta)}{2} \end{array} \right) \;=\; \psi|_{x=0} \;=\;
\lim_{\eta \searrow 0} e^{-\frac{L(\eta) - h(\eta)}{2}}
\left( \begin{array}{c} 1 \\ \displaystyle e^{-h(\eta)} \: 
\frac{\phi(\eta)}{2} \end{array} \right) . \]
We consider the two cases $\lim_{\eta \searrow 0} 
(L(\eta)-h(\eta))=0$ and $\neq 0$ separately. In the first 
case, the second components must have a non-zero limit (because 
$\psi(0) \neq 0$), and thus $\lim_{\eta \searrow 0} e^{-h(\pm 
\eta)} \phi(\pm \eta) = -\infty$. In the second case, the 
limits $\lim_{\eta \searrow 0} e^{-h(\pm \eta)}
\phi(\pm \eta)$ must exist and be equal. We conclude that
\begin{equation}
\lim_{\eta \rightarrow 0} e^{-h(-\eta)}\: \phi(-\eta) \;=\;
-\lambda \;=\; \lim_{\eta \rightarrow 0} e^{-h(\eta)}\:
\phi(\eta) \label{eq:7lambda}
\end{equation}
for some $\lambda \in [0, \infty]$. In
the case $\kappa_1=0$, this matching of the two ansatz' shows that
$\phi \leq 0$ for negative $x$, and thus we can make $\kappa_1$
slightly negative and repeat the above construction. Again 
using~(\ref{eq:7phi}), one deduces that $\lambda$ must
in fact be finite. We finally take the limit $\eta \rightarrow
0$ in~(\ref{eq:7d}) and~(\ref{eq:7e}) to obtain~(\ref{eq:72r}).
\QED

Our next goal is to bound $\vartheta(\xmin)$ uniformly in $\omega$. 
To this end, we combine the a-priori estimates for large $x$ 
(Lemmas~\ref{lemma71} and Lemma~\ref{lemma72}) with the estimates in a 
neighborhood of $x=0$ (Lemma~\ref{lemma73}). For negative $x$ outside 
of this neighborhood we can use similar methods as in Lemma~\ref{lemma62}.
\begin{Lemma} \label{lemma74}
There is $c>0$ such that for all $\omega \in (m-\delta, m)$,
\[ |\vartheta(\xmin)| \;\leq\; c \;. \]
\end{Lemma}
{\Proof} First let us verify that the assumptions of
Lemma~\ref{lemma73} are satisfied for a particular choice of
$\kappa_1$ and $\kappa_2$.  To this end, observe that $\phi(x)$ has no
zero for $x>0$, because otherwise
\[ \phi'(x) \;=\; 1 + h'(x)\: \sinh \phi \;=\; 1 \;, \]
violating the fact that $\phi'(x) \leq 0$ at the largest
zero (recall that Lemma~\ref{lemma71} implies that $\phi$ is negative 
for large $x$). Thus
\begin{equation}
\phi(x) \;<\; 0 \spc {\mbox{for all $x>0$.}}    \label{eq:7Z}
\end{equation}
As a consequence, $\sinh \phi < \phi$, and thus using~(\ref{eq:56}),
\[ \phi'(x) \;<\; 1 + h'(x)\: \phi \spc {\mbox{for all $x>0$.}} \]
Integrating this inequality from a given positive $x<x_0$ to $x_0$ and
using~(\ref{eq:7F}), we obtain the lower bound
\begin{equation}
\phi(x) \;>\; \underline{\phi}(x) \;=\; -\frac{1}{2}\: e^{h(x) -
h(x_0)} \:-\: e^{h(x)} \int_x^{x_0} e^{-h(\tau)}\:d\tau \spc
{\mbox{for $0<x<x_0$}} \label{eq:7Y}
\end{equation}
(this is indeed quite similar to the second part of~(\ref{eq:7e}),
but now we have solved for $\phi$ at the lower limit of the
integration range). According to~(\ref{eq:5nasy}), $\lim_{x
\rightarrow 0} \underline{\phi}(x) = 0$. We conclude that the
assumptions~(\ref{eq:7b}) are satisfied for $\kappa_1=0$ and
$\kappa_2>0$ sufficiently small.
We can further decrease $\kappa_1$ and increase $\kappa_2$,
provided that the bounds in~(\ref{eq:72r}) all take values in
the strip $(-\frac{1}{2}, 0)$.

The parameter $\lambda$ in~(\ref{eq:72r}) can be bounded a-priori. 
Namely, were $\lambda$ sufficiently large, we would get a
contradiction to~(\ref{eq:7Y}), whereas a very small value of
$\lambda$ would be inconsistent with~(\ref{eq:7E}).  Thus we can find
parameters
$0<\lambda_{\mbox{\scriptsize{min}}}<\lambda_{\mbox{\scriptsize{max}}}$
such that
\[ \lambda_{\mbox{\scriptsize{min}}} \;<\; \lambda \;<\;
\lambda_{\mbox{\scriptsize{max}}} \;. \]
As a consequence, in~(\ref{eq:72r}) the lower bound for
$\lambda=\lambda_{\mbox{\scriptsize{max}}}$ and the upper bound for
$\lambda=\lambda_{\mbox{\scriptsize{min}}}$ are a-priori bounds for
$\phi$.  We choose $x_2$ such that these bounds take values in the
strip $(-\frac{1}{2}, 0)$ on the interval $[x_2, 0)$.  Then we have
a-priori bounds for $\phi(x_2)$, and thus also for $\vartheta(x_2) =
\phi(x_2) - x_2$,
\begin{equation}
\vartheta_{\mbox{\scriptsize{min}}} \;<\; \vartheta(x_2) \;<\;
\vartheta_{\mbox{\scriptsize{max}}}\;.    \label{eq:7X}
\end{equation}

The bounds~(\ref{eq:7X}) are uniform in $\omega$. This is not 
surprising since the differential equation for $\vartheta$ involves 
only $h'$, which according to Lemma~\ref{lemma52} is bounded 
uniformly in $\omega$. To see this rigorously, one must be careful 
because $\lambda_{\mbox{\scriptsize{min}}}$ and
$\lambda_{\mbox{\scriptsize{max}}}$ {\em{do}} depend on $\omega$. 
Namely, according to~(\ref{eq:5nasy}), $h$ involves the additive 
constant $\frac{1}{3} \log \varepsilon^2$, which diverges as $\omega 
\nearrow m$. this implies, according to~(\ref{eq:7lambda}), that
\[ \varepsilon^{\frac{2}{3}}\: \vartheta_{\mbox{\scriptsize{min}}} 
\;,\spc \varepsilon^{\frac{2}{3}}\: \vartheta_{\mbox{\scriptsize{max}}} \]
can be chosen uniformly in $\varepsilon$. Using these scalings 
in~(\ref{eq:72r}), one sees that the estimates for $\vartheta$ and 
$x_2$ are indeed uniform.

It remains to control $\vartheta$ on the interval $[\xmin, x_2]$. 
According to~(\ref{eq:56}), there is $R>0$ independent of $\omega$
such that
\[ 1+h'(x) \: \sin(x+\vartheta) \;>\; \frac{1}{2} \spc {\mbox{for $x \in
[\xmin+R, x_2 - R]$}} \]
(note that this last interval is non-empty in view of
Lemma~\ref{lemma52a}).  On the bounded intervals $[\xmin, \xmin+R)$
and $(x_2-R, x_2]$ we can control $\vartheta$ directly by integrating
the equations in a method similar to~(\ref{eq:6NZ}).  In the
intermediate region, we integrate by parts and obtain similar
to~(\ref{eq:6NY}),
\[ |\vartheta(\xmin+R) - \vartheta(x_2-R)| \;\leq\; 2 (|h'(\xmin+R)| +
|h'(x_2-R)|) \:+\: 4 \int_{\sxmin+R}^{x_2-R} (|h''| + h'^2)\:d\tau
\;, \] and the terms on the right are all uniformly bounded
according to Lemma~\ref{lemma52}. \QED
The next lemma controls the $\omega$-dependence of $\vartheta$.
\begin{Lemma} \label{lemma75}
There is $c>0$ such that for all $\omega \in (m-\delta, m)$,
\[ |\vartheta_\omega(\xmin)| \;\leq\; \frac{c}{m-\omega}\;. \]
\end{Lemma}
{\Proof} In the proof of Lemma~\ref{lemma74}, we have verified that 
the hypothesis of Lemma~\ref{lemma73} are satisfied, and thus
$\vartheta(0)=0$ for all $\omega$. Hence
$\vartheta_\omega(\xmin)$ is obtained by integrating the
differential equation~(\ref{eq:3k}) from $\xmin$ to zero. This
gives in analogy to~(\ref{eq:12a}),
\[ \vartheta_\omega(\xmin) \;=\; e^{-L(\sxmin)} \int_{\sxmin}^0
e^{-L(\tau)}\:
h'_\omega(\tau)\: \sin \phi(\tau)\: d\tau\;. \]
By definition of $\Psi^{k \omega n}_1$, $\lim_{u \rightarrow -\infty}
L(u)=1$ (see \cite[eqn~(3.31)]{FKSY2} and~(\ref{eq:25a})).
Standard Gronwall estimates on the interval $(-\infty, \umin)$
show that $L(\xmin)$ is bounded uniformly in $\omega$.
Furthermore, it was shown in Lemma~\ref{lemma52} that
$(m-\omega)\:h_\omega$ has bounded total variation. Thus to
finish the proof, it suffices to show that there is $c$
independent of $\omega$ such that
\begin{equation}
\left| e^{-L(\tau)}\: \sin \phi(\tau) \right| \;\leq\; c \spc
{\mbox{for all $\tau \in [\xmin, 0)$}}. \label{eq:7T}
\end{equation}
The integration-by-parts technique of Lemma~\ref{lemma62} yields
that $L$ is uniformly bounded in the region $[\xmin, x_2]$ with
$x_2$ as in the proof of Lemma~\ref{lemma74} (for more details see
the last paragraph of Lemma~\ref{lemma74}, where this method is
used to estimate $\vartheta$). On the interval $(x_2, 0)$, the
a-priori bounds for $\phi$, (\ref{eq:72r}), show that
\begin{equation}
|\sin \phi(\tau)| \;\leq\; |\phi(\tau)| \;\leq\; c\: e^{h(\tau) -
h(x_2)} \label{eq:7V}
\end{equation}
(with $c$ independent of $\omega$). Furthermore,
\[ (h-L)' \;=\; h'\: (1-\cos \phi) \;\leq\; |h'|\: \phi^2 \;\leq\; c\:
x^{-\frac{1}{3}} \;, \] where in the last step we
used~(\ref{eq:7V}), (~\ref{eq:5nasy}), and~(\ref{eq:56}). Since
$x^{-\frac{1}{3}}$ is integrable,
\[ \left. (h - L) \right|_{x_2}^\tau \;\leq\; c \;. \]
We exponentiate and use that $L(x_2)$ is bounded to obtain
\begin{equation}
e^{-L(\tau)}\: e^{h(\tau) - h(x_2)} \;\leq\; c \;. \label{eq:7U}
\end{equation}
The inequality~(\ref{eq:7T}) follows by combining~(\ref{eq:7V})
and~(\ref{eq:7U}). \QED

\section{Proof of the Decay Rates}
\setcounter{equation}{0}
We are now ready to finish the proof of Theorem~\ref{thm1}. In
view of Theorem~\ref{thm21} and the consideration in the last
paragraph of Section~\ref{sec6}, it remains to show that the
contribution to the propagator~(\ref{eq:1a}) for $\omega \in
(m-\delta, m)$ has the decay~(\ref{eq:1decay}). Since the
coefficients $t_{ab}$ are trivial for $\omega<m$, (\ref{eq:k}), 
the contribution to the propagator~(\ref{eq:1b}) simplifies to
\begin{equation}
\Psi(t,x) \;=\; \frac{1}{\pi} \: \sum_{|k| \leq k_0} \:\sum_{|n|
\leq n_0} \int_{m-\delta}^m d\omega \:e^{-i \omega t} \left[
\sum_{a,b=1}^2 \Psi^{k \omega n}_1(x) \:\bra
\Psi^{k \omega n}_1 \:|\: \Psi_0 \ket \right] \;.    \label{eq:8A}
\end{equation}
Since $\Psi_0$ has compact support, it
suffices to analyze the $\omega$-dependence of $\Psi^{k \omega
n}_1(u)$ for $u$ in a compact set.

According to the separation ansatz~(\ref{eq:323}), we must only analyze the
radial function $X$ (the angular part $Y$ is clearly smooth in $\omega$).
To see the $\omega$-dependence of $X$ in detail, we
substitute~(\ref{eq:25a})
into~(\ref{eq:2phi}) and~(\ref{eq:19a}). This gives, exactly as in the case
$|\omega|>m$, the formula~(\ref{eq:3e}). For fixed $u$, the function
$h$ in~(\ref{eq:3e}) depends smoothly on $\omega$. Using that $h$ vanishes
at
the event horizon (because $\lim_{r \searrow r_1} f(r) = \omega = \lim_{r
\searrow r_1} g(r)$ according to~(\ref{eq:23a}) and
(\ref{eq:2xx}),(\ref{eq:2w})), our normalization condition for $\Psi^{k
\omega
n}_1$ near the event horizon~(\ref{eq:2norm}) yields that
\[ 1 \;=\; \lim_{u \rightarrow -\infty} |X(u)|^2 \;=\; 2\: \lim_{u
\rightarrow
-\infty} e^{-L(u)} \]
and thus $\lim_{u \rightarrow -\infty} L(u) = \log 2$, independent of
$\omega$. Furthermore, an argument similar to~(\ref{eq:6hp}) shows that
$h'(u)$ has exponential decay as $u \rightarrow -\infty$. Hence standard
Gronwall estimates yield that $L(u)$ is bounded and depends smoothly on
$\omega$. Furthermore, Gronwall estimates in the finite region between
$\umin$ and $u$ show that the difference $\vartheta(u) - \vartheta(\umin)$
is
uniformly bounded and smooth in $\omega$. Writing
\[ (x+\vartheta)(u) \;=\; (x+\vartheta)(\umin) \:+\:
((x(u) - x(\umin)) + (\vartheta(u) - \vartheta(\umin))) \;, \] we
conclude that the only possible non-smooth terms in~(\ref{eq:3e}) are
the factors $\cos (\pmin/2)$ and $\sin(\pmin/2)$ with $\pmin \equiv \xmin +
\vartheta(\xmin)$.

We next consider the factors $\bra \Psi^{k \omega
n}_1 \:|\: \Psi_0 \ket$ in~(\ref{eq:8A}).  Again from Gronwall
estimates, one sees that for $\omega>m$, the expectation values $\bra
\Psi^{k \omega n}_2 \:|\: \Psi_0 \ket$ depend smoothly on $\omega$,
and thus our assumption~(\ref{eq:1as1}) implies that
\[ r_2 \;\equiv\; \lim_{\omega \searrow m} \bra
\Psi^{k \omega n}_2 \:|\: \Psi_0 \ket \;\neq\; 0 \;. \]
Except for the additional phase factors, the
expectation values are smooth even for $\omega<m$. To compute the
phases, we consider~(\ref{eq:3e}) in the asymptotic regime $u \rightarrow
-\infty$, and compare with~(\ref{eq:c}) and~(\ref{eq:d}). This
shows that for $\omega \in (m-\delta, m)$,
\begin{equation}
\bra \Psi^{k \omega n}_1 \:|\: \Psi_0 \ket \;=\; r_1\:
\alpha_1\:
\exp \left( -i \:\frac{\pmin}{2} \right) \:-\: r_2\: \alpha_2\:
\exp \left( i \:\frac{\pmin}{2} \right)
\label{eq:8B}
\end{equation}
with coefficients $\alpha_a$ which depend smoothly on
$\omega$ and are non-zero (indeed $\lim_{\omega \nearrow m} \alpha_a =1$). 
Since the factor $r_2$ is non-zero, we conclude that $\bra \Psi^{k
\omega n}_1 \:|\: \Psi_0 \ket$ has a non-vanishing contribution which
oscillates like $\exp (i \pmin/2)$.  Using~(\ref{eq:8B}) and~(\ref{eq:3e})
in~(\ref{eq:8A}), we can write the propagator in the region $\omega \in
(m-\delta, m)$ as the Fourier integral
\[ \int_{m-\delta}^m e^{-i \omega t} \left( s_1 \: e^{-i (\sxmin +
\vartheta(\sxmin))} \:+\: s_2 \:+\: s_3 \: e^{i (\sxmin +
\vartheta(\sxmin))} \right) d\omega \]
with coefficients $s_j$ which are smooth in $\omega$ and $s_3
\neq 0$.  According to Lemma~\ref{lemma52a}, Lemma~\ref{lemma74}, and
Lemma~\ref{lemma75}, the three contributions to this Fourier integral
satisfy the hypotheses of Lemma~\ref{lemma41}, Lemma~\ref{lemma40},
and Lemma~\ref{lemma42}, respectively (with $\alpha = m-\omega$). 
Hence the first two terms decay like $t^{-\frac{5}{6}-\varepsilon}$,
whereas the last term gives the desired decay rate $\sim
t^{-\frac{5}{6}}$.  This concludes the proof of Theorem~\ref{thm1}.

\section{Probability Estimates}
\label{sec9}
\setcounter{equation}{0}
We now proceed with the proof of Theorem~\ref{thm2}. We want to 
compute the probability $p$, (\ref{eq:1pdef}). We begin with the following
lemma.
\begin{Lemma} \label{lemma91}
For any Schwartz function $f \in {\cal{S}}(\R \times \R)$, let $A_\pm$ be
defined by
\[ A_\pm \;=\; \lim_{t \rightarrow \infty} \int_{-\infty}^{u_0} du
\int_{-\infty}^\infty
d\omega \int_{-\infty}^\infty d\omega' \:e^{-i (\omega - \omega')
(t \pm u)} \: f(\omega, \omega') \;. \]
Then
\begin{equation}
A_+ \;=\; 2 \pi \int_{-\infty}^\infty f(\omega,
\omega)\: d\omega \spc{\mbox{and}}\spc A_- \;=\; 0 \;.
\label{eq:9A}
\end{equation}
\end{Lemma}
{\Proof}
We integrate by parts to obtain
\begin{eqnarray*}
\lefteqn{
\int_{-\infty}^\infty d\omega \int_{-\infty}^\infty d\omega'
\:e^{-i (\omega - \omega')(t \pm u)} \: f(\omega, \omega') } \\
&=& \int_{-\infty}^\infty d\omega \int_{-\infty}^\infty d\omega' \:
\frac{1}{(t \pm u)^2 + 1} \left( (\partial_\omega+1)(\partial_{\omega'}+1)
\:e^{-i (\omega - \omega')(t \pm u)} \right) f(\omega, \omega') \\
&=& \int_{-\infty}^\infty d\omega \int_{-\infty}^\infty d\omega' \:
\frac{1}{(t \pm u)^2 + 1}\: e^{-i (\omega - \omega')(t \pm u)} \:
g(\omega, \omega') \;,
\end{eqnarray*}
where $g$ is the Schwartz function
\begin{equation}
g(\omega, \omega') \;=\; (-\partial_\omega+1)(-\partial_{\omega'}+1)
f(\omega, \omega') \;. \label{eq:9D}
\end{equation}
Since the factor $((t \pm u)^2 + 1)^{-1}$ is integrable in $u$, we can
integrate over $u$, apply Fubini, and use Lebesgue's dominated
convergence theorem to take the limit $t \rightarrow \infty$ inside
the integrand,
\begin{eqnarray}
A_\pm &=& \lim_{t \rightarrow \infty} \int_{-\infty}^{u_0} du
\int_{-\infty}^\infty d\omega
\int_{-\infty}^\infty d\omega'
\frac{1}{(t \pm u)^2 + 1}\:e^{-i (\omega - \omega')(t \pm u)} \: g(\omega,
\omega') \nonumber \\ &=& \int_{-\infty}^\infty d\omega
\int_{-\infty}^\infty
d\omega'
\: g(\omega, \omega') \;
\lim_{t \rightarrow \infty} \int_{-\infty}^{u_0}
\frac{1}{(t \pm u)^2 + 1}\:
e^{-i (\omega - \omega')(t \pm u)}\: du \label{eq:9C}
\end{eqnarray}

In the case ``$t-u$,'' we introduce a new integration variable
$\alpha = t-u$ and get for the inner integral
\[ \lim_{t \rightarrow \infty} \int_{-\infty}^{u_0} du
\frac{1}{(t \pm u)^2 + 1}\:e^{-i (\omega - \omega')(t-u)} \;=\;
\lim_{t \rightarrow \infty} \int_{t-u_0}^\infty 
\frac{1}{\alpha^2+1}\:e^{-i (\omega - \omega')\alpha} \: d\alpha \;=\; 0 \;.
\] This proves that $A_-=0$.

In the case ``$t+u$,'' we obtain similarly an integral over the real
line, which can be computed with residues,
\[ \lim_{t \rightarrow \infty} \int_{-\infty}^{u_0} du
\frac{1}{(t \pm u)^2 + 1}\:e^{-i (\omega - \omega')(t+u)} \;=\;
\int_{-\infty}^\infty \frac{1}{\alpha^2+1} \:e^{-i (\omega - \omega')\alpha}
\: d\alpha
\;=\; \pi \: e^{-|\omega-\omega'|} \;. \]
We substitute this formula as well as~(\ref{eq:9D}) into~(\ref{eq:9C}) and
integrate by parts ``backwards'',
\begin{eqnarray}
A_+ &=& \int_{-\infty}^\infty d\omega \int_{-\infty}^\infty d\omega'
\; g(\omega, \omega')\; \pi\: e^{-|\omega-\omega'|} \nonumber \\
&=& \pi \int_{-\infty}^\infty d\omega \int_{-\infty}^\infty d\omega'
f(\omega, \omega') \left( (\partial_\omega+1)(\partial_{\omega'}+1)
e^{-|\omega-\omega'|} \right) \label{eq:9E}
\end{eqnarray}
A short explicit calculation shows that the derivatives can be computed
in the distributional sense to be
\[ (\partial_\omega+1)(\partial_{\omega'}+1)
e^{-|\omega-\omega'|} \;=\; 2 \: \delta(\omega-\omega') \;. \]
Substitution into~(\ref{eq:9E}) gives the desired formula for $A_+$.
\QED
We remark that the above lemma cannot be obtained by naively
interchanging the orders of integration.

\begin{Thm} \label{thm91}
The probability $q$ for the Dirac particle to disappear into the event
horizon, defined for any $\varepsilon>0$ by
\begin{equation}
q \;=\; \lim_{t \rightarrow \infty} \int_{\{r_1 < r <
r_1+\varepsilon \}} (\overline{\Psi} \gamma^j \Psi)(t,x) \:\nu_j
\; d\mu \;, \label{eq:9Z}
\end{equation}
is given by
\begin{equation}
q \;=\; \frac{1}{\pi} \: \sum_{|k| \leq k_0} \:\sum_{|n| \leq n_0}
\int_{-\infty}^\infty d\omega \sum_{a,b=1}^2 s^{k \omega n}_{ab}
\bra \Psi^{k \omega n}_b \:|\: \Psi_0 \ket
\bra \Psi_0 \:|\: \Psi^{k \omega n}_a \ket \label{eq:9I}
\end{equation}
with
\begin{equation}
s^{k \omega n}_{ab} \;=\; \left\{ \begin{array}{cl} \delta_{a1}\:
\delta_{b1} & {\mbox{if $|\omega| \leq m$}} \\
2\: t^{k \omega n}_{2a} \:\overline{t^{k \omega n}_{2b}}
& {\mbox{if $|\omega| > m$.}}
\end{array} \right.  \label{eq:9J}
\end{equation}
\end{Thm}
We remark that $p+q=1$, since we know from~\cite{FKSY2} that the probability
for the Dirac particle to be in any compact set tends to zero as $t
\rightarrow \infty$. \\[.3em]
{\em{Proof of Theorem~\ref{thm91}. }}
In the variable $u$, we need to compute the probability for the Dirac
particle
to be in the region $u<u_0$, where $u_0$ may be chosen as small as we like. 
Thus we can work with the asymptotic formulas near the event horizon,
with error terms which decay exponentially fast as $u_0 \rightarrow
-\infty$.  More precisely, a straightforward calculation shows that
the probability integral in~(\ref{eq:9Z}) coincides asymptotically
with the integral of the scalar product $\bra .|.  \ket$ on the
transformed spinors (see~\cite[eqn~(2.15)]{FKSY2}). Thus it
suffices to consider the probability
\begin{equation}
q(t) \;=\; \int_{-\infty}^{u_0} du \int_{-1}^1 d\cos \vartheta
\int_0^{2 \pi} d\varphi\: \bra \Psi \:|\: \Psi \ket_{|(t, u,
\vartheta, \varphi)} \;, \label{eq:90}
\end{equation}
and let $t \rightarrow \infty$. Due to the additivity of the
probabilities corresponding to the angular momentum modes (which are
orthogonal with respect to the scalar product $\bra .|. \ket$),
it suffices to consider a solution of the Dirac equation with fixed angular
momentum quantum numbers
$k$ and $n$, i.e.
\[ \Psi(t,x) \;=\; \frac{1}{\pi} \: \int_{-\infty}^\infty d\omega
\:e^{-i \omega t} \sum_{a,b=1}^2 t^\omega_{ab}\:\Psi^{k \omega
n}_a(x) \:\bra \Psi^{k \omega n}_b \:|\: \Psi_0 \ket  \;. \]
We substitute this formula for the propagator into~(\ref{eq:90}) and carry
out the angular integrals to obtain
\begin{eqnarray*}
q(t) &=& \frac{1}{\pi^2} \:\int_{-\infty}^{u_0} du
\int_{-\infty}^\infty d\omega \int_{-\infty}^\infty d\omega' \:
e^{-i (\omega-\omega') t} \\
&&\spc\times \: \sum_{a,b,c,d=1}^2
t^\omega_{ab}\: \bra \Psi^\omega \:|\: \Psi_0 \ket \;
\overline{t^{\omega'}_{cd}\: \bra \Psi^{\omega'} \:|\: \Psi_0 \ket} \:
\bra X^{\omega'}_d \:|\: X^\omega_a \ket(u) \;.
\end{eqnarray*}
Substituting for $X^\omega$ the asymptotic formulas~(\ref{eq:36}), valid
near
the event horizon, we obtain with an exponentially small error term
\begin{eqnarray}
\lefteqn{ q(t) \;=\; \frac{1}{\pi^2} \:\int_{-\infty}^{u_0} du
\int_{-\infty}^\infty d\omega \int_{-\infty}^\infty d\omega' \:
e^{-i (\omega-\omega') t} \:\sum_{a,b,c,d=1}^2 } \nonumber \\
&&\times \: t^\omega_{ab}\: \bra \Psi^\omega_b \:|\: \Psi_0 \ket \;
\overline{t^{\omega'}_{cd}\: \bra \Psi^{\omega'}_d \:|\: \Psi_0 \ket} \:
\left( \overline{f^{\omega'}_{c+}} f^w_{a+}\: e^{-i(\omega-\omega')u}
\:+\: \overline{f^{\omega'}_{c-}} f^w_{a-}\: e^{i(\omega-\omega')u}
\right) . \label{eq:9Y}
\end{eqnarray}
Since we cannot expect the integrand to be smooth when
$\omega$ or $\omega'$ is equal to $\pm m$, we must use an approximation
argument. Namely, the integrand is bounded and
has rapid decay in $\omega$ and $\omega'$. Thus we can approximate the
integrand in $L^1$ by a Schwartz function, and applying
Lemma~\ref{lemma91} we obtain
\begin{equation}
q \;=\; \frac{1}{\pi} \int_{-\infty}^\infty d\omega\:
\sum_{a,b,c,d=1}^2 
2 \:t^\omega_{ab}\:\overline{t^{\omega}_{cd}}\:
\overline{f^\omega_{c+}} f^\omega_{a+}\; 
\bra \Psi^\omega_b \:|\: \Psi_0 \ket \; \overline{
\bra \Psi^\omega_d \:|\: \Psi_0 \ket} \;. \label{eq:9H}
\end{equation}

It remains to compute the factors $\overline{f^\omega_{c+}} f^\omega_{a+}$.
In the case $|\omega| > m$, we conclude from~(\ref{eq:d}) that
\begin{equation}
\overline{f^\omega_{c+}} f^\omega_{a+} \;=\; \delta_{c1}\: \delta_{a1}\;.
\spc {\mbox{(for $|\omega| \geq m$).}} \label{eq:9F}
\end{equation}
On the other hand if $|\omega| \leq m$, using~(\ref{eq:k}) in~(\ref{eq:9H})
we must only compute $|f^\omega_{1+}|^2$. To this end, we again
consider~(\ref{eq:3e}). Using that $h$ vanishes asymptotically near the
event
horizon, one sees that
\[ \lim_{r \searrow r_1} (|X^+|^2 - |X_-|^2) \;=\; 0 \;, \]
and thus $|f^\omega_{1+}|^2 = |f^\omega_{1-}|^2$. Furthermore, our
normalization of the fundamental solutions near the event
horizon~(\ref{eq:2norm}) yields that
$|f^\omega_{1+}|^2 + |f^\omega_{1-}|^2 = 1$, and thus we conclude that
\begin{equation}
|f^\omega_{1+}|^2 \;=\; \frac{1}{2} \spc {\mbox{(for $|\omega|<m$).}}
\label{eq:9G}
\end{equation}
Substituting~(\ref{eq:9F}) and~(\ref{eq:9G}) into~(\ref{eq:9H}) and
using~(\ref{eq:k}) completes the proof.
\QED
{\em{Proof of Theorem~\ref{thm2}. }} 
Since the initial data is normalized by $\bra \Psi_0 | \Psi_0 \ket=1$,
by taking the inner product of $\Psi_0$ with~(\ref{eq:1b}), evaluated at
$t=0$, we obtain that
\begin{equation}
1 \;=\; \frac{1}{\pi} \: \sum_{|k| \leq k_0} \:\sum_{|n| \leq n_0}
\int_{-\infty}^\infty d\omega \sum_{a,b=1}^2 t^{k \omega n}_{ab}
\bra \Psi^{k \omega n}_b \:|\: \Psi_0 \ket
\bra \Psi_0 \:|\: \Psi^{k \omega n}_a \ket \;. \label{eq:9X}
\end{equation}
As remarked after the statement of Theorem~\ref{thm91}, $p=1-q$. Thus $p$ is
obtained by taking the difference of~(\ref{eq:9X}) and~(\ref{eq:9I}).
Using~(\ref{eq:k}), we get~(\ref{eq:1prob}).

For the proofs of~{\bf{(i)}}--{\bf{(iv)}}, it again suffices to consider a
fixed angular momentum mode. Since the energy distribution in the interval
$[-m,m]$ is absent from~(\ref{eq:1prob}), it is obvious that~{\bf{(iii)}}
holds.

To prove~{\bf{(ii)}}, we introduce a vector $v^\omega \in \C^2$ by
\[ v^\omega_a \;=\; \bra \Psi^\omega_a \:|\: \Psi_0 \ket \;,\spc
a=1,2 \]
and remark that in the region $|\omega|>m$ we can write the integrands
in~(\ref{eq:9X}) and~(\ref{eq:1prob}) as
\[ \langle T^\omega\: v^\omega \:|\: v^\omega \rangle \spc{\mbox{and}}\spc
\langle A^\omega\: v^\omega \:|\: v^\omega \rangle \;, \]
respectively, where, using Lemma~\ref{lemma31},
\[ T^\omega \;=\; \left( \begin{array}{cc} 1/2 & t^\omega_{12} \\
\overline{t^\omega_{12}} & 1/2 \end{array} \right)
\spc{\mbox{and}}\spc
A^\omega \;=\; \left( \begin{array}{cc} 0 & 0 \\ 0 &
1/2 - 2 \:|t^\omega_{12}|^2
\end{array} \right). \]
An easy calculation shows that $T \geq A$. Thus from~(\ref{eq:9X})
and~(\ref{eq:k}),
\begin{eqnarray}
1 &=& \frac{1}{\pi} \int_{-m}^m d\omega \: |\bra \Psi^\omega_1 \:|\:
\Psi^\omega_1 \ket|^2 \:+\: \frac{1}{\pi} \int_{\sR \setminus [-m,m]}
d\omega \: \langle T^\omega\: v^\omega \:|\: v^\omega \rangle
\label{eq:9V} \\
&\geq& \frac{1}{\pi} \int_{-m}^m d\omega \: |\bra \Psi^\omega_1 \:|\:
\Psi^\omega_1 \ket|^2 \:+\: \frac{1}{\pi} \int_{\sR \setminus [-m,m]}
d\omega \: \langle A^\omega\: v^\omega \:|\: v^\omega \rangle
\label{eq:9T} \\
&=& \frac{1}{\pi} \int_{-m}^m d\omega \: |\bra \Psi^\omega_1 \:|\:
\Psi^\omega_1 \ket|^2 \:+\: p \;, \label{eq:9U}
\end{eqnarray}
and this is strictly larger than $p$ because in case~{\bf{(ii)}} the first
summand is positive.

To prove~{\bf{(i)}}, we note that the factor $|\bra \Psi^\omega_2 \:|\:
\Psi_0 \ket|^2$ is positive on a set of positive measure (by continuity in
$\omega$). Thus it suffices to show that
\[ \frac{1}{2} - 2\: |t^\omega_{12}|^2 \;>\; 0 \spc {\mbox{for all
$\omega \in \R \setminus [-m,m]$}}. \]
Using the explicit formula~(\ref{eq:13b}) in Lemma~\ref{lemma31}, this holds
iff
\begin{equation}
|\tanh z| \;<\; 1 \label{eq:9W}
\end{equation}
with $z$ as in~(\ref{eq:6zf}). Using~(\ref{eq:2y}) together
with~(\ref{eq:3d}), we see that $-2\pi < \vartheta^{(1)}_\infty -
\vartheta^{(2)}_\infty < 2 \pi$ (by the uniqueness theorem for ODEs).
Then from~(\ref{eq:6zf}),
\[ -\frac{\pi}{4} \;<\; \arg z \;<\; \frac{\pi}{4}\;. \]
It follows that $|e^{2z}-1| < |e^{2z}+1|$, giving~(\ref{eq:9W}). This
proves~{\bf{(i)}}.

Finally, if~(\ref{eq:1as1}) holds, then we saw in~(\ref{eq:8B}) that $\bra
\Psi^\omega_1 \:|\: \Psi_0 \ket$ is non-zero for $\omega \in (m-\delta, m)$.
Thus~{\bf{(iv)}} follows from~{\bf{(i)}} and~{\bf{(ii)}}.
\QED
Given the fact that he Fourier transform of a $C^{\infty}$ function
with compact support is analytic, one might think that
$\bra \Psi^{k \omega n}_b \:|\: \Psi_0 \ket$ should be analytic in $\omega$,
implying that the cases~{\bf{(ii)}} and~{\bf{(iii)}} cannot occur.
However, it is not at all obvious that the solutions of our ODEs should
depend analytically on $\omega$. Should this be the case,
one could still make sense of~{\bf{(ii)}} and~{\bf{(iii)}}
by slightly weakening the assumptions on the initial data.

We conclude by describing the class of initial data for which the Dirac
particle must escape to infinity, with probability one.
\begin{Corollary} \label{cor91}
The probability $p$ is equal to one if and only if the initial data
satisfies
for all $k$, $\omega$, and $n$ the following conditions,
\[ \left\{ \begin{array}{cl}
\bra \Psi^{k \omega n}_1 \:|\: \Psi_0 \ket \;=\; 0 &
{\mbox{if $|\omega| \leq m$}} \\
\bra \Psi^{k \omega n}_1 \:|\: \Psi_0 \ket \;=\; -2\:t^{k \omega n}_{12}\:
\bra \Psi^{k \omega n}_2 \:|\: \Psi_0 \ket &
{\mbox{if $|\omega| > m$}}. \end{array} \right. \]
\end{Corollary}
{\Proof}
It again suffices to consider a fixed angular momentum mode.
In view of~(\ref{eq:9U}), $p=1$ only if
\begin{equation}
\int_{-m}^m d\omega \: |\bra \Psi^\omega_1 \:|\:
\Psi^\omega_1 \ket|^2 \;=\; 0 \;, \label{eq:9M}
\end{equation}
and thus the energy distribution of the initial data must be supported in
the
outside the interval $(-m,m)$. Furthermore, the inequality in~(\ref{eq:9T})
must be replaced by equality, and thus
\begin{equation}
\langle S^\omega\: v^\omega \:|\: v^\omega \rangle \;=\; 0
\spc {\mbox{for all $\omega \in \R \setminus [-m,m]$}}, \label{eq:9S}
\end{equation}
where the matrix $S^\omega$ is defined by
\[ S^\omega \;=\; T^\omega-A^\omega
\;=\; \left( \begin{array}{cc} 1/2 & t^\omega_{12} \\
\overline{t^\omega_{12}} & 2 \: |t^\omega_{12}|^2 \end{array} \right) . \]
The eigenvalues of $S^\omega$ are zero and $\frac{1}{2} + 2
|t^\omega_{12}|^2
>0$. Hence~(\ref{eq:9S}) implies that $v^\omega$ must be in the kernel of
$S^\omega$, i.e.
\begin{equation}
v^\omega_1 \;=\; -2 \:t^\omega_{12} \:v^\omega_2 \;. \label{eq:9N}
\end{equation}
Conversely, if~(\ref{eq:9M}) and~(\ref{eq:9N}) hold, it is obvious
from~(\ref{eq:9V})--(\ref{eq:9U}) that
$p=1$.
\QED
One can also understand directly why $p=1$ for special choices
of the initial data. Indeed, to obtain such initial data, one can consider
the physical situation where a Dirac particle at time $t=-\infty$ comes in
from spatial infinity. If we take as our initial data the corresponding
$\Psi(t,x)$ at $t=0$ and reverse the direction of time, the solution to this
Cauchy problem will clearly escape to infinity with probability one.\\[.7em]
{\em{Acknowledgments:}} We would like to thank Paul Koosis and Robert Wald
for helpful discussions. Furthermore, we want to thank the referee for helpful
suggestions and comments.

\begin{tabular}{lcl}
\\
Felix Finster & $\;\;\;\;$ & Niky Kamran\\
Max Planck Institute for && Department of Math.\ and Statistics \\
Mathematics in the Sciences && McGill University \\
Inselstr.\ 22-26 && Montr{\'e}al, Qu{\'e}bec \\
04103 Leipzig, Germany && Canada H3A 2K6  \\
Felix.Finster@mis.mpg.de && nkamran@math.McGill.CA \\
\\
Joel Smoller & $\;\;$ & Shing-Tung Yau \\
Mathematics Department && Mathematics Department \\
The University of Michigan && Harvard University \\
Ann Arbor, MI 48109, USA && Cambridge, MA 02138, USA \\
smoller@umich.edu && yau@math.harvard.edu
\end{tabular}

\end{document}